\documentclass{article}

\usepackage[OT1]{fontenc}
\usepackage{alignpred_arxiv,times}

\usepackage{amsmath}
\usepackage{amssymb}
\usepackage{bm}

\usepackage{xspace}

\newcommand{\figtexttight}{\vspace{-0.8\baselineskip}}
\newcommand{\wraptabtightabove}{\vspace{-1\baselineskip}}
\newcommand{\wrapfigtopnudge}{%
  \vspace{-\dimexpr1.15\intextsep\relax}%
  \vspace{-0.35ex}%
}
\newcommand{\wrapfigcaptiongap}{\vspace{0.28\baselineskip}}
\newcommand{\wrapfigbelowgap}{\vspace{1.85\baselineskip}}

\usepackage{booktabs}
\usepackage{algorithm}
\usepackage{algorithmic}
\usepackage{graphicx}
\usepackage{wrapfig}
\usepackage[section,above]{placeins}

\makeatletter
\def\fps@figure{tbp}
\def\fps@table{tbp}
\makeatother
\usepackage{xcolor}
\usepackage[hidelinks]{hyperref}
\hypersetup{pdftitle={AlignPred: Error-Controlled KV Cache Compression via Alignment and Predictive Coding},
pdfauthor={Xu Yang, Jiapeng Zhang, Zhiwei Zuo, Yuxin Chen, Feiqiang Sun, Chengguang Xu, Feng Jin, Zhuo Tang}}
\usepackage{url}
\title{\raggedright AlignPred: Error-Controlled KV Cache\\
Compression via Alignment and Predictive Coding}

\author{\textbf{Xu Yang\textsuperscript{1}\thanks{These authors contributed equally.}\thanks{Email: \texttt{yangxv@hnu.edu.cn}},\enspace
Jiapeng Zhang\textsuperscript{1}\footnotemark[1],\enspace
Zhiwei Zuo\textsuperscript{3},\enspace Yuxin Chen\textsuperscript{2}}\\
\textbf{Feiqiang Sun\textsuperscript{2},\enspace Chengguang Xu\textsuperscript{2},\enspace
Feng Jin\textsuperscript{2},\enspace Zhuo Tang\textsuperscript{1}\thanks{Corresponding author.}}\\
\textnormal{\textsuperscript{1}Hunan University\qquad\textsuperscript{2}Tencent\qquad
\textsuperscript{3}Central China Normal University}}

\iclrfinalcopy 
\date{}

\begin{document}

\raggedbottom

\maketitle
\thispagestyle{preprintfirst}

\begin{abstract}
Long-running agents retain KV caches across turns and sessions for hours or
days. Compressing these persistent caches requires reducing storage while
controlling reconstruction error. We present AlignPred, a codec that combines
local alignment with predictive coding under a configurable KV
reconstruction-error budget. Local
orthogonal transforms align related heads or layers into a common coordinate
frame, allowing AlignPred to combine token-wise prediction with same-token
prediction across heads or layers. Residuals are quantized on an
error-bounded lattice and entropy-coded, without learned predictors,
encoder-side mode search, or large joint transforms. Across three attention
mechanisms, AlignPred requires less storage than the evaluated baselines at
comparable measured reconstruction error. Against the evaluated CPU codec
implementations, it decodes at least $1.6\times$ faster while matching or
improving encode time.
AlignPred achieves near-native task scores at substantially compressed
operating points on LongBench-Pro at 128K context. AlignPred's SGLang HiCache integration reduces
persistent target-KV storage by 42--72\% relative
to native HiCache, with comparable latency.
\end{abstract}

\setlength{\parskip}{4.7pt}
\setlength{\textfloatsep}{14pt plus 2pt minus 4pt}
\setlength{\floatsep}{10pt plus 2pt minus 2pt}
\section{Introduction}
\label{sec:introduction}

\begin{wrapfigure}[11]{r}{0.39\textwidth}
    \centering
    \setlength{\columnsep}{16pt}%
    \setlength{\intextsep}{8pt}%
    \wrapfigtopnudge
    \includegraphics[width=0.90\linewidth]{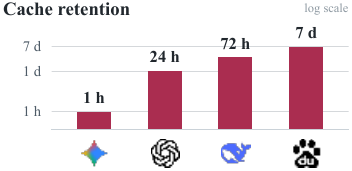}
    \wrapfigcaptiongap
    \caption{Long-lived KV caches motivate compact storage.}
    \label{fig:api-cache-retention}
    \wrapfigbelowgap
\end{wrapfigure}
Long-running agents repeatedly revisit previously processed instructions,
documents, and tool outputs across multiple
turns~\citep{zheng2024sglang,abhyankar2024infercept}. Retaining the corresponding
KV caches avoids recomputing the same prefixes when a session
resumes~\citep{kwon2023pagedattention,liu2024cachegen}, but makes otherwise
transient KV caches persistent. Commercial APIs already support retaining KV
caches for hours or even days (Figure~\ref{fig:api-cache-retention}). As both context length and the
number of retained sessions grow, compact storage becomes increasingly
important for agent serving.

Serving frameworks such as TensorRT-LLM~\citep{nvidia2026trtllmkv} and
vLLM~\citep{vllm2026fp8kv} support FP8 KV-cache quantization, which shows that
models can tolerate some numerical error in their KV caches. The acceptable
amount of error, however, depends on the application. Fidelity-sensitive
workloads require tight error bounds, while other workloads can accept greater
approximation. KV compression should therefore let users set a
reconstruction-error budget and maximize the compression ratio within that
budget. Existing KV compression methods typically seek high compression ratios
while preserving performance on a limited set of downstream
tasks~\citep{liu2024kivi,hooper2024kvquant,staniszewski2026kvtc}, without
explicitly bounding KV reconstruction error. As a result, users cannot tell
whether a given setting meets the accuracy requirements of their own
applications.

Meeting an error budget is only part of the problem. Effective compression
still depends on how the resulting information is represented and coded.
Current methods still leave room to exploit the properties of KV caches.
General-purpose error-bounded compressors such as SZ3~\citep{liang2022sz3}
ignore head and layer geometry.
Error-bounded coding has reached KV caches in the GPU-resident
setting~\citep{jiang2025kvcomp,jiang2026packkv}, where decompression fuses
into every attention step. \citet{jiang2025kvcomp} found that a prediction
stage helps compression but conflicts with that fusion, and dropped it.
Persistent KV storage reconstructs the cache once at restore time, so
prediction remains available.
KVTC~\citep{staniszewski2026kvtc} uses alignment to build a joint transform
under a bit budget, not prediction references.
Persistent storage thus leaves prediction available, but no existing method
uses alignment to supply prediction references. This raises a natural question.
Can alignment enable simple prediction in the original KV layout?

AlignPred starts from KV geometry. Within one attention head or
latent-cache block, vectors from different tokens share a common
coordinate frame, making token-wise dependencies directly accessible.
Related heads or layers may encode similar information in different
coordinate frames, so their corresponding coordinates are not directly
comparable. We apply local orthogonal rotations to align these coordinate frames.
After alignment, each cache value has two complementary sources of
redundancy, neighboring tokens within the same block and the same token
across related blocks. Alignment makes this cross-block relation
usable as a prediction reference.
Figure~\ref{fig:coding-intuition} shows this structure.

We exploit this structure with alignment-conditioned predictive coding.
Because alignment exposes the cross-block dependency, prediction requires
only fixed arithmetic. The first block is reconstructed by token
interpolation, and later blocks are predicted from previously reconstructed
aligned blocks before error-bounded residual coding. Alignment uses only one
small rotation per block, sized to its head or latent dimension, so it costs
far less computation than the joint transforms used in prior methods. This
design also avoids learned predictors and encoder-side mode search while
preserving parallelism across independent cache objects.

\paragraph{Contributions.}
\textbf{(1)} We show that local alignment enables effective cross-block
prediction: in ablations on all three models, the same predictor increases
storage without alignment but reduces it after alignment.
\textbf{(2)} We introduce AlignPred, a codec that exploits this structure
through fixed token-wise and cross-block prediction followed by error-bounded
residual coding. It preserves the cache's block structure and supports a
configurable reconstruction-error budget. Across three attention mechanisms,
it stores less than the evaluated baselines at comparable measured error and
decodes at least $1.6\times$ faster than the evaluated same-host CPU
implementations.
\textbf{(3)} We integrate the codec into SGLang HiCache, reducing persistent
target-KV storage by 42--72\% versus native HiCache at comparable cache-hit
latency.

\section{Background and Motivation}
\label{sec:related}

\begin{figure}[tb]
    \centering
    \includegraphics[width=\linewidth]{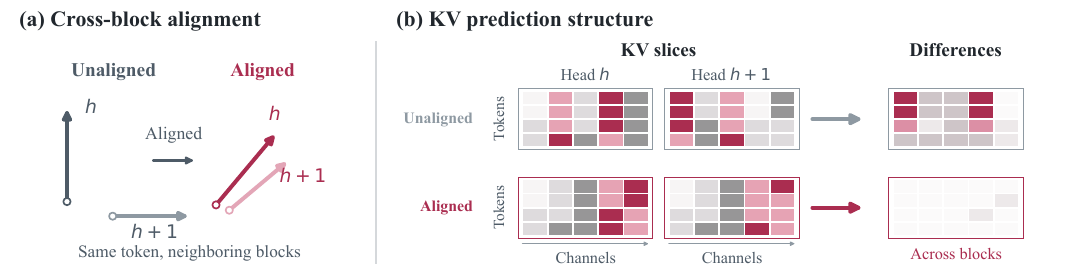}
    \caption{Alignment of related KV blocks.
    (a) Rotation registers same-token vectors across heads or layers.
    (b) Token rows share channels. Alignment also reduces cross-block
    differences.}
    \label{fig:coding-intuition}
    \figtexttight
\end{figure}

\paragraph{Error-controlled compression.}
Relative reconstruction error normalizes the reconstruction error by the
magnitude of the original cache.
Error-controlled compression treats a bound on this quantity as the
constraint~\citep{di2016sz,tao2017sz,liang2018sz2}. Rate-controlled compression instead fixes a rate, such as a
bit width or bit budget, and then minimizes distortion under that rate
or accepts the distortion it induces~\citep{goyal2001transformcoding}.
We therefore bound relative reconstruction error and report storage
against the measured reconstruction error.

\paragraph{Predictive coding.}
Predictive compression represents each value using a prediction and a
quantized residual~\citep{elias1955predictive,ibarria2003lorenzo}. Accurate prediction concentrates the residual
distribution and reduces entropy-coded storage. Encoder and decoder form
predictions from the same reconstructed references to keep their states
synchronized, while residual symbols are entropy-coded losslessly.
This motivates prediction rules tailored to KV cache structure.

\begin{figure}[tb]
    \centering
    \includegraphics[width=\linewidth]{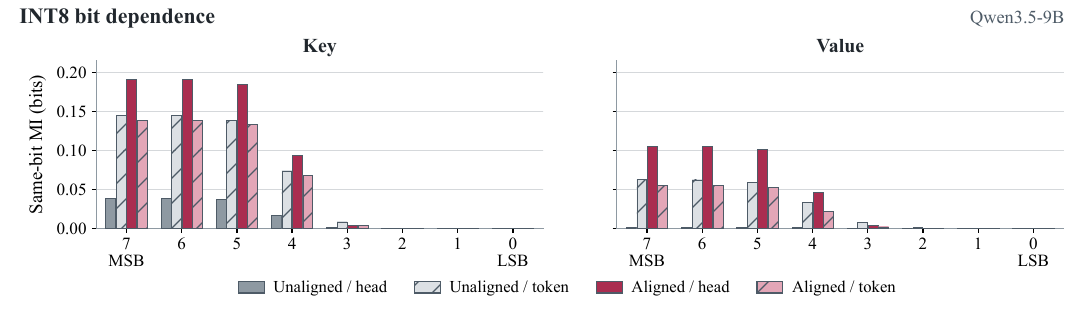}
    \caption{After alignment, high-order-bit MI is higher across neighboring
    heads than across neighboring tokens on average in \mbox{Qwen3.5-9B}.}
    \label{fig:bit-dependence}
    \figtexttight
\end{figure}

\paragraph{KV dependencies for prediction.}
Figure~\ref{fig:coding-intuition} shows that local orthogonal rotations
can align the coordinate frames of related KV blocks.
Figure~\ref{fig:bit-dependence} measures what this alignment exposes at
the bit level. After centering, caches are uniformly quantized to signed
8-bit codes, with or without local alignment. Same-bit mutual
information (MI) is computed on those diagnostic INT8 codes on
Qwen3.5-9B, comparing the same token across neighboring heads with
adjacent tokens in the same head. After alignment, high-order bits of
the same token across heads have greater MI than adjacent tokens in the
same head, on average for Keys and Values. Alignment also increases
this cross-head MI relative to unaligned blocks. After alignment, the
same-token cross-head relation can therefore serve as a prediction
reference, and at high-order bits it can even be stronger than
neighboring-token dependence. Coupled Quantization~\citep{zhang2024coupledquantization}
exploits cross-channel dependence through joint quantization, while
KVTC~\citep{staniszewski2026kvtc} uses aligned geometry for joint transforms.
Here we compare neighboring-token and aligned cross-head dependencies
at the bit level as prediction references. The same
neighboring-block construction applies to grouped-query
heads~\citep{ainslie2023gqa} and to layers of a shared latent
cache~\citep{deepseekai2024deepseekv2,brandon2024cla,liu2024minicache}.

\paragraph{Persistent KV storage.}
Our setting is persistent KV storage, which differs from both remote
streaming and GPU-resident KV compression.
Serving systems such as CachedAttention~\citep{gao2024cachedattention},
Mooncake~\citep{qin2025mooncake}, and LMCache~\citep{liu2025lmcache} keep
reusable KV caches in CPU memory, SSDs, or remote storage, so cache size
sets their capacity and transfer cost.
KVTC~\citep{staniszewski2026kvtc} combines PCA-based transform coding,
quantization, and entropy coding. KVCodec~\citep{mi2026kvcodec} encodes quantized KV with a video
codec. Scientific compressors such as SZ3~\citep{liang2022sz3},
cuSZ~\citep{tian2020cusz}, and ZFP~\citep{lindstrom2014zfp} code generic
floating-point arrays. CacheGen~\citep{liu2024cachegen} and LLM.265~\citep{xu2025llm265}
target streaming or hardware paths. Palu~\citep{chang2025palu},
MatryoshkaKV~\citep{lin2025matryoshkakv}, and xKV~\citep{chang2026xkv}
share aligned geometry but reduce rank rather than predict residuals.

\section{Method}
\label{sec:method}

\setcounter{topnumber}{1}
\setcounter{bottomnumber}{0}
\setcounter{totalnumber}{1}

AlignPred compresses KV caches under a reconstruction-error budget. It locally
aligns related blocks, applies token-wise and cross-block prediction, and
error-bounds and entropy-codes the residuals. Decoding reverses these steps,
restoring the original layout.

\begin{figure}[t]
    \centering
    \includegraphics[width=\linewidth]{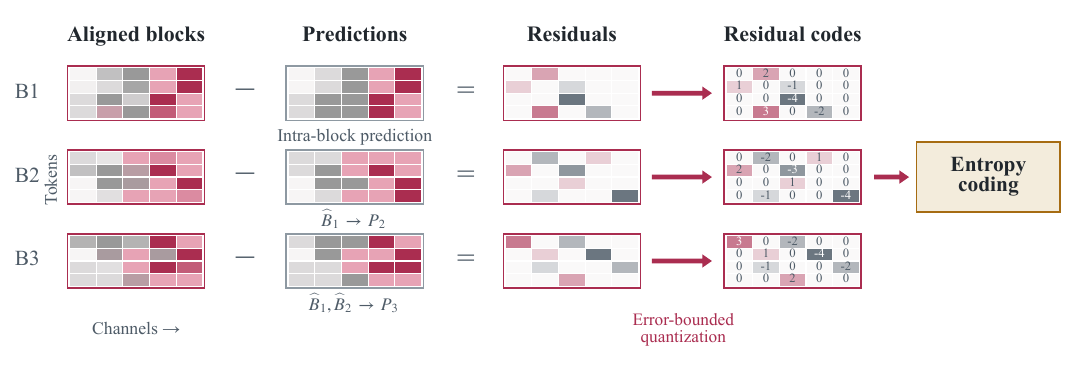}
    \caption{Predictive residual coding for three aligned cache blocks: token-wise prediction initializes the first block, and cross-block prediction uses reconstructed blocks to predict subsequent blocks. Residuals undergo error-bounded quantization and lossless entropy coding.}
    \label{fig:predictive-residual-coding}
    \figtexttight
\end{figure}

\subsection{Problem Formulation}
\label{sec:method-formulation}

Let $X$ denote one independently coded KV cache object. Given a user-specified error
target $\epsilon$, the encoder produces a bitstream $B$, from which the
decoder reconstructs $\widehat X$. Error-controlled compression uses a
bound on relative reconstruction error as the design objective,
\begin{equation}
    \min\;\operatorname{size}(B)
    \qquad\mathrm{subject\ to}\qquad
    \frac{\|X-\widehat X\|_F}{\|X\|_F}\leq\epsilon.
    \label{eq:method-exact-codes}
\end{equation}
Section~\ref{sec:method-quantization} maps $\epsilon$ to a residual step
that attains this bound in exact arithmetic. Storage comparisons use
measured reconstruction error. Encoding and decoding
efficiency are additional design objectives.

\subsection{Local Coordinate Alignment}
\label{sec:method-gpa}
\label{sec:method-calibration-data}

Each cache block contains cached vectors at a set of positions from one
attention head or one layer of a shared latent
cache~\citep{deepseekai2024deepseekv2}. Inter-block
dependencies are therefore dependencies across heads or layers,
depending on the model. We align related
blocks in small groups. The index $v$ identifies the corresponding head or
layer within a group. Calibration samples $S_v\in\mathbb{R}^{n_c\times d}$ are used to estimate both the mean
$\mu_v$ and the orthogonal transform $A_v$. Samples from different heads or layers in
the same group correspond to the same cache positions.

We first estimate the mean from the $n_c$ calibration samples:
\begin{equation}
    \mu_v=\frac{1}{n_c}S_v^\top\mathbf{1}.
    \label{eq:method-calibration-centering}
\end{equation}
After subtracting $\mu_v$ and normalizing each row, we obtain $\bar S_v$.
For a group of $m$ blocks, GPA~\citep{gower1975gpa} fits the rotations by
aligning these calibration samples to a common representation $M$:
\begin{equation}
    \min_{\{A_v\},M}\sum_{v=1}^{m}\|\bar S_vA_v-M\|_F^2,
    \qquad A_v^\top A_v=I,
    \label{eq:method-gpa-objective}
\end{equation}
where $A_v\in\mathbb{R}^{d\times d}$ and
$M\in\mathbb{R}^{n_c\times d}$. The fitted mean and rotation are shared
across cache blocks associated with the same head or layer.

For a cache block $X_v\in\mathbb{R}^{N\times d}$ to be compressed, we apply
these calibrated parameters directly:
\begin{equation}
    Z_v=(X_v-\mathbf{1}\mu_v^\top)A_v.
    \label{eq:method-analysis-transform}
\end{equation}
The aligned values $Z_v$ are used for prediction. Row normalization is only
used during fitting. Here $\mathbf{1}$ is an all-ones vector of the required
length.

\paragraph{Calibration data.}
We support two calibration options: \textbf{(1) auxiliary calibration}
learns rotations and means from auxiliary KV caches for reuse.
\textbf{(2) Cache-specific calibration} avoids concerns about out-of-domain
transfer by fitting directly to the target cache, with one parameter bank reused
across its chunks. This option incurs per-cache fitting and parameter storage.
We use auxiliary calibration by default to amortize these costs across caches.
Section~\ref{sec:robustness} evaluates the trade-off.

\subsection{Intra- and Inter-Block Prediction}
\label{sec:method-prediction}

Figure~\ref{fig:predictive-residual-coding} shows the prediction and
coding process for three aligned blocks.
Let $z_{v,t,c}$ denote an aligned value in block $v$, token position $t$,
and channel $c$, where $v=1,\ldots,m$ and $t=0,\ldots,N-1$.
Its prediction is $p_{v,t,c}$. Adding the dequantized residual gives
$\widehat z_{v,t,c}$, which both encoder and decoder then use as a later
reference, always within the same channel.

\paragraph{Intra-block prediction.}
We compute a channel mean for the current group of blocks:
\begin{equation}
    b_c=\frac{1}{mN}\sum_{v=1}^{m}\sum_{t=0}^{N-1}z_{v,t,c}.
    \label{eq:method-prediction-mean}
\end{equation}
This stored mean summarizes the current group, rather than the calibration
samples used for $\mu_v$. In the first block, even token positions are
predicted from $b_c$ and reconstructed first. Odd positions then use the
average of their reconstructed neighbors:
\begin{equation}
    p_{1,t,c}=
    \begin{cases}
        b_c, & t\text{ even},\\
        (\widehat z_{1,t-1,c}+\widehat z_{1,t+1,c})/2,
            & t\text{ odd}.
    \end{cases}
    \label{eq:method-token-prediction}
\end{equation}
This establishes the first reconstructed block.

\paragraph{Inter-block prediction.}
Block $v\geq2$ averages earlier reconstructions at the same token and
channel, then blends that average with $b_c$:
\begin{equation}
    a_{v,t,c}=\frac{1}{v-1}\sum_{j=1}^{v-1}\widehat z_{j,t,c},
    \qquad
    p_{v,t,c}=b_c+\lambda_v(a_{v,t,c}-b_c).
    \label{eq:method-cross-block-prediction}
\end{equation}
We set $\lambda_2=0.5$ and $\lambda_v=0.75$ for $v\geq3$.

\subsection{Error-Controlled Coding and Reconstruction}
\label{sec:method-quantization}

\begin{figure}[t]
    \centering
    \includegraphics[width=\linewidth]{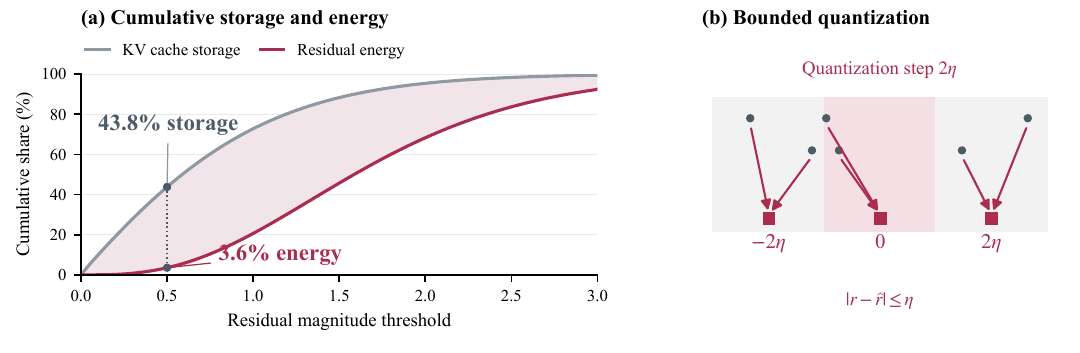}
    \caption{Error-bounded residual quantization. (a) Small residuals occupy 43.8\% of KV cache storage but contribute only 3.6\% of residual energy. (b) Quantization merges nearby values within the error bound.}
    \label{fig:residual-storage-error}
    \figtexttight
\end{figure}

Figure~\ref{fig:residual-storage-error}(a) shows the imbalance between
storage cost and residual error contribution, motivating precision
reduction under an explicit reconstruction-error target.

\paragraph{Error target.}
Given a target relative reconstruction error $\epsilon$, the encoder
sets a residual step for an object $X$ containing $n$ scalar values:
\begin{equation}
    \eta=\epsilon\frac{\|X\|_F}{\sqrt n}.
    \label{eq:method-relative-error}
\end{equation}
The step is computed once per object, without reconstruction search.
Because orthogonal rotations preserve the error norm, the per-value check
$|z_i-\widehat z_i|\leq\eta$ implies
$\|X-\widehat X\|_F\leq\sqrt n\,\eta$, and hence relative error at most
$\epsilon$, in exact arithmetic. Reported reconstruction errors are measured
after inverse transformation and conversion to the source format, including
the effects of finite-precision arithmetic.

\paragraph{Residual quantization.}
For each aligned value $z_i$ and prediction $p_i$, we quantize the residual
$r_i=z_i-p_i$ on a grid with spacing $2\eta$. Figure~\ref{fig:residual-storage-error}(b)
shows that residuals within each quantization
interval map to the same symbol for entropy coding. Let $\widehat r_i$
denote the dequantized residual. The reconstructed value satisfies
\begin{equation}
    \widehat z_i=p_i+\widehat r_i,
    \qquad |z_i-\widehat z_i|\leq\eta.
    \label{eq:method-residual-error}
\end{equation}
The encoder checks this bound and stores $z_i$ directly if the check fails
or the residual exceeds the coding range. Reconstructed references, spacing
$2\eta$, and an escape for values that miss the bound are the standard
error-bounded residual lattice~\citep{liang2022sz3}. After local alignment we
do not search Lorenzo versus interpolation on the raw array. We use the
fixed KV schedule of Section~\ref{sec:method-prediction}.

\paragraph{Encoding and reconstruction.}
We losslessly compress the quantization codes using asymmetric numeral
systems (ANS)~\citep{duda2013ans}. Decoding recovers the codes and follows the same prediction order
to reconstruct $\widehat Z_v$. We then restore the original coordinates as
$\widehat X_v=\widehat Z_vA_v^\top+\mathbf{1}\mu_v^\top$
and convert to the source format. Appendix~\ref{sec:quantization-implementation}
gives the object codec as Algorithm~\ref{alg:predictive-kv-codec}.

\section{System Integration}
\label{sec:system}

\begin{figure}[tb]
    \centering
    \includegraphics[width=\linewidth]{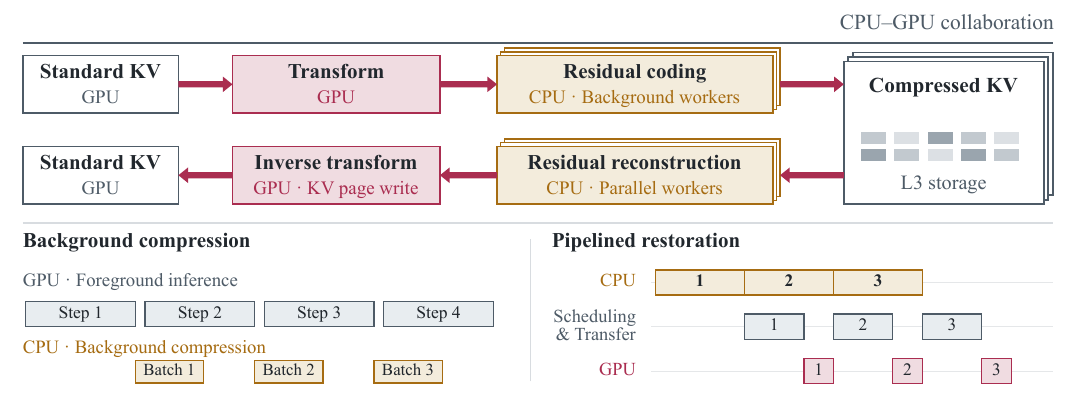}
    \caption{Overlap of background KV compression with foreground inference,
    and pipelined CPU--GPU restoration. Each CPU compression job starts after
    its corresponding GPU step.
    Timeline lengths are schematic and illustrate overlap rather than measured
    stage durations.
    L3 denotes HiCache's persistent file tier.}
    \label{fig:system-integration}
    \figtexttight
\end{figure}

\paragraph{CPU--GPU cooperation.}
Figure~\ref{fig:system-integration} shows how compression and restoration
overlap with inference. After the GPU produces a cache block, CPU workers
compress it in the background. On a cache hit, the CPU reconstructs the
aligned cache while the GPU converts decoded groups back to the model's native
layout. Qwen uses HiCache's layerwise restoration path. For GLM and DeepSeek,
rank 0 decodes each shared-latent group and broadcasts it to the
tensor-parallel ranks. Double buffering overlaps CPU decoding, transfer, and
GPU restoration. DeepSeek writes the result through its native FP8 cache path.
Further implementation details are given in Appendix~\ref{sec:appendix-system}.

\paragraph{Inverse transforms.}
Serving restores native coordinates with an explicit inverse rotation per
cached token. When the attention path permits, the rotation and mean can be
absorbed into subsequent projections (Appendix~\ref{sec:appendix-gqa}). Our
measurements do not use this path.
\FloatBarrier
\setcounter{topnumber}{3}
\setcounter{bottomnumber}{2}
\setcounter{totalnumber}{4}

\section{Experiments}
\label{sec:experiments}

\subsection{Experimental setup}
\label{sec:exp-setup}

We evaluate three attention mechanisms.
Qwen3.5-9B stores GQA Keys and Values, GLM-5.3-Flash stores the post-LN
MLA latent, and DeepSeek-V4-Flash stores the $4\times$ CSA pool and already
uses FP8 as the default KV cache.
GLM RoPE, indexer, and KDA states are outside the compressed object.
DeepSeek storage savings are measured relative to its already pooled cache.

Unless a later subsection states otherwise, means and rotations are fitted
on 8,192 uniformly sampled positions from each of five disjoint calibration
documents covering English, Chinese, code, mathematics, and retrieval, and
are then frozen. Captured caches are converted to FP16.
Storage is bits per value (bpv), including codec metadata and excluding
reusable calibration parameters.

Offline experiments compare adapted coding strategies on common caches
using storage size, reconstruction error, and same-host CPU cost.
LongBench-Pro evaluates task quality. Ablations examine design choices.
Serving experiments compare
AlignPred with native HiCache storage.

\subsection{Compression at Comparable Reconstruction Error}
\label{sec:numerical-rd}

\begin{wraptable}{r}{0.5\textwidth}
\wraptabtightabove
\caption{Cost of the evaluated CPU codec implementations relative to Ours
at the FP8-near operating points in Table~\ref{tab:matched-storage}, range
over the three models. Above $1$ favors Ours.}
\label{tab:matched-cost-summary}
\vspace{0.5\baselineskip}
\begingroup
\footnotesize
\setlength{\tabcolsep}{2pt}
\renewcommand{\arraystretch}{1.0}
\begin{tabular*}{\linewidth}{@{\extracolsep{\fill}}lccc@{}}
\toprule
Baseline & Storage & Encode & Decode \\
\midrule
KVCodec & 1.10--1.30 & 5.3--6.8 & 5.4--6.3 \\
KVTC-style & 1.13--1.18 & 1.3--1.6 & 3.3--3.7 \\
Rotation + SZ3 & 1.07--1.10 & 1.02--1.04 & 1.6--2.1 \\
\bottomrule
\end{tabular*}
\endgroup

\vspace{-1.5\baselineskip}
\end{wraptable}

\begin{figure}[tb]
    \centering
    \includegraphics[width=\linewidth]{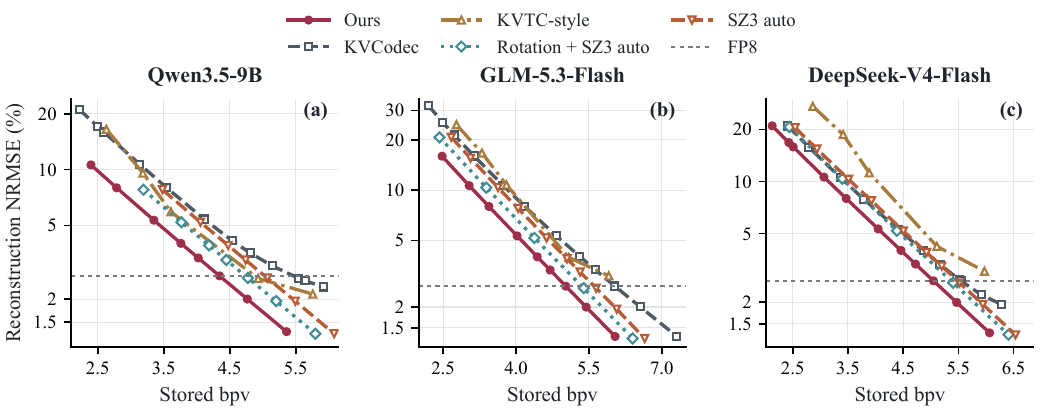}
    \caption{Storage size versus reconstruction error on five InfiniteBench
    categories, including KVCodec, KVTC-style, Rotation + SZ3 auto, and unrotated
    SZ3 auto. Error axes are logarithmic. Dotted lines mark FP8 error.
    Axis ranges vary by model.}
    \label{fig:measured-rd}
    \figtexttight
\end{figure}

\paragraph{Setup.}
We evaluate storage versus reconstruction error on
InfiniteBench~\citep{zhang2024infinitebench}.
Error is the relative Frobenius norm over the evaluation set.
Figure~\ref{fig:measured-rd} plots storage size against reconstruction error.

\paragraph{Baselines.}
We compare coding strategies on identical cache representations at comparable
measured reconstruction error, using a KVTC-style adaptation for the
common-input setting.
Our KVCodec~\citep{mi2026kvcodec} reference uses software HEVC on a token
layout. Our KVTC-style~\citep{staniszewski2026kvtc} reference uses joint PCA and
quantization. SZ3 auto~\citep{liang2022sz3} runs on native or locally
aligned caches with the same frozen means.
Inputs retain positional encoding where applicable. Implementation
settings are in Appendix~\ref{sec:measured-metrics}.

\paragraph{Results.}
Across all three mechanisms, Ours stores less than the evaluated KVCodec,
KVTC-style, Rotation + SZ3, and unrotated SZ3 implementations at comparable
measured error (Figure~\ref{fig:measured-rd}).
Unrotated SZ3 is strictly more expensive than Rotation + SZ3 at matched
error, so local alignment also reduces the rate of a generic error-bounded
compressor.
With alignment held fixed, this comparison shows the advantage of our
coding pipeline over SZ3 auto. Table~\ref{tab:component-measured}
separately evaluates the contribution of inter-block prediction.
Table~\ref{tab:matched-cost-summary} summarizes storage and same-host CPU
coding cost at the operating points in Table~\ref{tab:matched-storage}.
Ours decodes faster than all three timed CPU implementations, encodes faster
than the evaluated KVCodec and KVTC-style instances, and ties Rotation + SZ3 on encode.
At the FP8 reconstruction-error level, Ours uses 36.9--45.7\% less
storage than 8-bit FP8.
These savings persist for the lifetime of each retained cache.
Additional results are in Appendix~\ref{sec:measured-domain-bpv}.

\subsection{Full-context downstream quality}
\label{sec:task-quality}

\begin{figure}[tb]
    \centering
    \includegraphics[width=\linewidth]{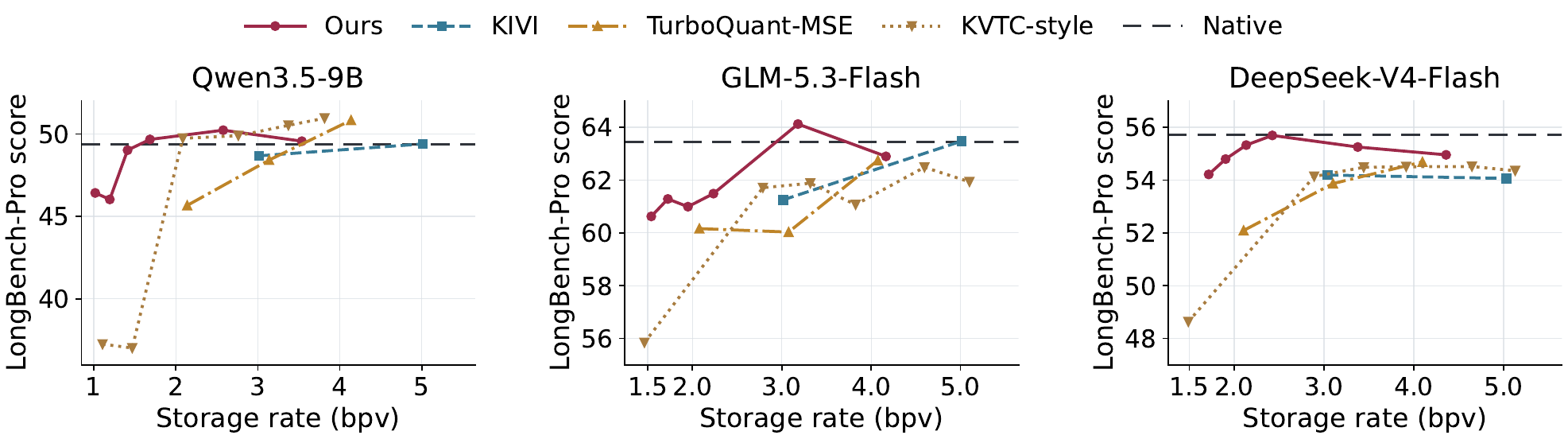}
    \caption{LongBench-Pro task score versus storage rate (bpv).
    KIVI and TurboQuant-MSE provide GPU-resident quantization references.}
    \label{fig:task-quality}
    \figtexttight
\end{figure}

\paragraph{Setup.}
This experiment evaluates how downstream quality changes as persistent KV
storage is reduced.
We measure LongBench-Pro task score~\citep{chen2026longbenchpro} across
compression operating points in the 128K setting, using the
official scoring protocol (Appendix~\ref{sec:quality-measurements}).
Decoding uses non-thinking prompts, greedy search, and at most
1,024 generated tokens.
Our method reuses the frozen auxiliary bank from
Section~\ref{sec:exp-setup} throughout this evaluation.
Figure~\ref{fig:task-quality} reports task score against storage rate in bpv.

\paragraph{Baselines.}
Persistent-storage codecs reuse the settings of
Section~\ref{sec:numerical-rd}.
KIVI and TurboQuant-MSE~\citep{zandieh2025turboquant} provide
reference points for task quality at different storage sizes using
GPU-resident KV quantization.

\paragraph{Results.}
Figure~\ref{fig:task-quality} shows that Ours preserves near-native
LongBench-Pro quality over a wide range of storage rates on Qwen, and at
several more compressed operating points on GLM and DeepSeek.
Across the evaluated operating points, lower reconstruction error is generally
associated with higher downstream quality, while local score fluctuations make
the relationship imperfectly monotonic. This makes reconstruction error a
practical control variable for selecting compression strength, with downstream
evaluation used to determine the acceptable range for each workload.
Additional results are in Appendix~\ref{sec:quality-measurements}.

\subsection{Ablation studies}
\label{sec:component-ablation}

\begin{figure}[tb]
    \centering
    \begin{minipage}[t]{0.49\textwidth}
        \vspace{0pt}
        \makeatletter\def\@captype{table}\makeatother
        \caption{Component ablation on one English cache per model (bpv $\downarrow$).}
        \label{tab:component-measured}
        \vspace{4pt}
        \centering\small
\begingroup
\setlength{\tabcolsep}{1.2pt}
\renewcommand{\arraystretch}{1.03}
\begin{tabular*}{\linewidth}{@{\extracolsep{\fill}}ccccrrr@{}}
\toprule
\multicolumn{4}{c}{Components enabled} & \multicolumn{3}{c}{Stored bpv $\downarrow$} \\
\cmidrule(r){1-4}\cmidrule(l){5-7}
Alignment & \multicolumn{2}{c}{Prediction} & $\lambda_v$ & Qwen & GLM & DeepSeek \\
\cmidrule(lr){2-3}
 & Inter & Intra & & & & \\
\midrule
$\times$ & $\times$ & $\times$ & --- & 4.973 & 5.626 & 5.662 \\
$\times$ & $\times$ & $\checkmark$ & --- & 4.755 & 5.364 & 5.341 \\
$\checkmark$ & $\times$ & $\checkmark$ & --- & 4.698 & 5.316 & 5.325 \\
$\times$ & $\checkmark$ & $\checkmark$ & $\checkmark$ & 5.052 & 5.585 & 5.544 \\
$\checkmark$ & $\checkmark$ & $\times$ & $\checkmark$ & 4.561 & 5.186 & 5.289 \\
$\checkmark$ & $\checkmark$ & $\checkmark$ & $\times$ & 4.581 & 5.234 & 5.357 \\
\midrule
$\checkmark$ & $\checkmark$ & $\checkmark$ & $\checkmark$ & \textbf{4.524} & \textbf{5.168} & \textbf{5.281} \\
\bottomrule
\end{tabular*}
\endgroup

    \end{minipage}\hfill
    \begin{minipage}[t]{0.49\textwidth}
        \vspace{0pt}
        \centering
        \makeatletter\def\@captype{figure}\makeatother
        \caption{Calibration ablation. Hatched region is
        cache-specific parameter storage.}
        \label{fig:calibration-strategies}
        \vspace{4pt}
        \includegraphics[width=\linewidth]{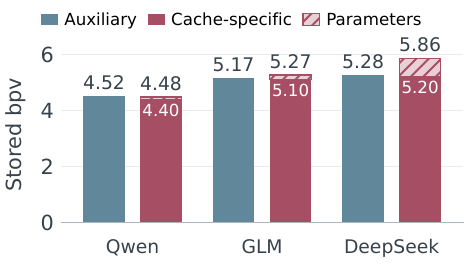}
    \end{minipage}
\end{figure}

\paragraph{Setup.}
We use ablations to examine how intra-block prediction, local alignment,
inter-block prediction, and blending interact, and whether cache-specific
alignment is worth its fitting and parameter cost.
Both ablations use one English cache per model at matched reconstruction
error to isolate component effects and calibration overhead.
Table~\ref{tab:component-measured} toggles alignment, inter-block
prediction, intra-block prediction, and the default blending weights.
Figure~\ref{fig:calibration-strategies} compares the default auxiliary
calibration with cache-specific calibration.

\paragraph{Component results.}
Local alignment makes inter-block prediction effective in all three
evaluated models. Under the default blend, adding inter-block prediction
to intra-block prediction increases storage by 0.203--0.297~bpv without
alignment, but reduces storage by 0.044--0.174~bpv after alignment
(Table~\ref{tab:component-measured}). Replacing unaligned intra-block
prediction with aligned inter-block prediction using the default blend,
with intra-block prediction disabled, saves 0.194, 0.178, and 0.052~bpv
on Qwen, GLM, and DeepSeek, respectively.

Intra-block prediction alone saves 0.218--0.321~bpv over the no-prediction
reference, which uses error-lattice quantization and rANS.
Setting $\lambda_v=1$, which removes shrinkage toward the group mean,
increases storage by 0.057--0.076~bpv relative to the full configuration.
The full configuration has the lowest storage in this ablation for all three
models, saving 0.381--0.458~bpv over the no-prediction reference.

\paragraph{Calibration results.}
\label{sec:robustness}
Cache-specific calibration reduces the compressed payload, but its
parameter overhead offsets the savings on two of the three evaluated
models. Figure~\ref{fig:calibration-strategies} shows smaller payloads
on all three models. After including the per-cache parameter bank, the
net benefit remains positive on Qwen at the evaluated length but is
offset on GLM and DeepSeek. These results support auxiliary calibration
as the default trade-off at the evaluated lengths. Longer caches amortize cache-specific parameter
storage over more cached values, whereas repeated reuse amortizes the
one-time fitting cost.
Additional results are in Appendix~\ref{sec:measured-domain-bpv}.

\FloatBarrier
\subsection{End-to-end serving efficiency}
\label{sec:serving-efficiency}
\label{sec:online-write}

\begin{figure}[tb]
    \centering
    \includegraphics[width=\linewidth]{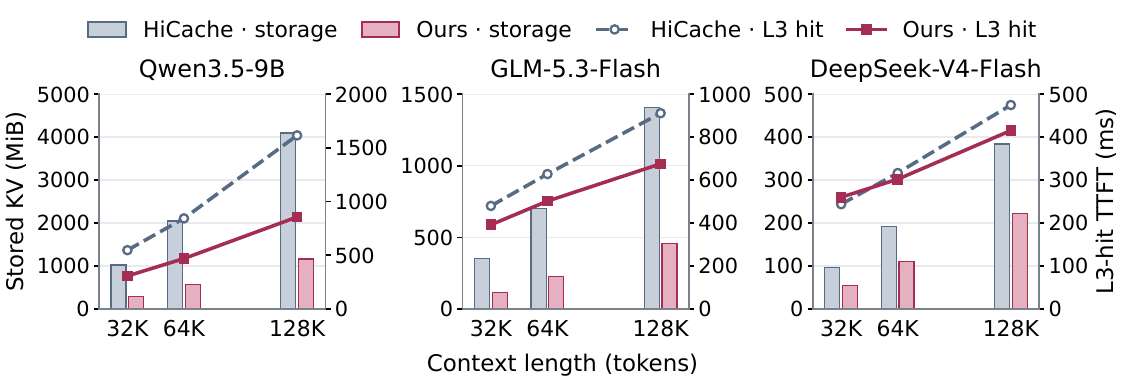}
    \caption{Persistent target-KV file size and L3-hit restore TTFT versus
    native HiCache, after a restart with empty L1/L2. Batch size 1.
    Axis ranges vary by model.}
    \label{fig:serving-storage-ttft}
    \figtexttight
\end{figure}

\paragraph{Setup.}
This experiment evaluates persistent KV storage savings and cache-hit
restore latency when AlignPred is integrated into SGLang HiCache, using
the operating setting selected near the FP8 reference error in the offline
experiments.
We run a write-then-rehit roundtrip at 32K, 64K, and 128K with
$B=1$, 2, 4, and 8 concurrent requests.
The first requests populate L3 while prefill is in progress.
After a restart with empty L1/L2, the same requests restore their prefixes
from L3.
Figure~\ref{fig:serving-storage-ttft} reports target-KV file size and
L3-hit restore TTFT at $B=1$.
Further details and the concurrent results are provided in
Appendix~\ref{sec:appendix-system}.

\paragraph{Results.}
AlignPred reduces persistent target-KV storage by 42--72\% versus native
HiCache, including on DeepSeek, whose native cache is already FP8
(Figure~\ref{fig:serving-storage-ttft}).
The smaller payload also shortens cache-hit restore: on Qwen and GLM,
L3-hit TTFT is lower at every context length and stays lower under most
concurrent loads.
DeepSeek stays close to native for a single request but falls behind under
concurrency, because its restore runs synchronously on the scheduler thread.
Online compression overlaps with prefill and adds little to first-request
TTFT.

\section{Conclusion}

We present AlignPred, a reconstruction-error-controlled alignment-and-prediction
codec for persistent KV caches.
Local orthogonal alignment turns cross-head and cross-layer geometric
similarity into coordinate-wise predictive redundancy, enabling lightweight
cross-block prediction while preserving the original KV layout.
More broadly, our results suggest a design principle for persistent KV cache
compression: align related KV blocks to expose predictable structure, then
code the resulting residuals under an explicit error budget.

\setlength{\parskip}{6pt}
\setlength{\textfloatsep}{20pt plus 2pt minus 4pt}
\setlength{\floatsep}{12pt plus 2pt minus 2pt}
\clearpage
\subsection*{AI use statement}

In this work, we used generative AI tools to polish writing, to find related
work, and to assist with research execution by implementing experiment code
following the authors' designs. The core hypotheses and method are the
authors' own; generative AI tools were not used for research ideation. We have
reviewed all AI-assisted work and take responsibility for the final content of
this work, including text, claims or artifacts produced with the aid of
generative AI.

\begingroup
\setlength{\bibsep}{4pt}
\bibliography{references}
\bibliographystyle{alignpred}
\endgroup

\appendix
\setlength{\parskip}{5pt}
\setlength{\textfloatsep}{14pt plus 2pt minus 4pt}
\setlength{\floatsep}{10pt plus 2pt minus 2pt}
\section{Motivation for Error-Controlled KV Compression}
\label{sec:appendix-motivation}

Deploying lossy KV compression requires both numerical control and task-level
validation. Benchmark accuracy alone does not provide a transferable numerical
constraint for unseen workloads, making it difficult to reason about how
aggressively a codec can be configured. AlignPred instead exposes an explicit
reconstruction-error target, providing a workload-independent numerical
fidelity constraint for selecting and comparing operating points. This bound
does not guarantee downstream task quality; rather, it narrows the numerical
design space and makes application-specific validation more systematic.
Task-level checks remain necessary to determine which error targets are
acceptable for a given workload.

\begin{figure}[tbp]
    \centering
    \includegraphics[width=\linewidth]{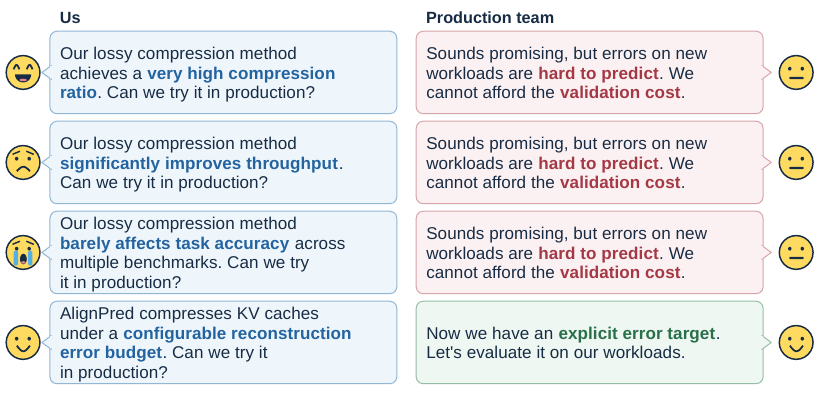}
    \caption{Illustrative dialogue motivating error-controlled KV compression.
    A configurable reconstruction-error budget supports different application
    requirements and provides a numerical reference for deployment validation.}
    \label{fig:motivation-dialogue}
\end{figure}

\FloatBarrier
\section{Codec Specification and Analysis}
\label{sec:appendix-method-details}

\subsection{End-to-end codec specification}
\label{sec:quantization-implementation}
\label{sec:codec-specification}
Algorithm~\ref{alg:predictive-kv-codec} is the object codec used in all
experiments. Fitted means and rotations come from
Section~\ref{sec:gpa-fitting-details}. Geometry and packet fields are in
Section~\ref{sec:frame-implementation}.

\begin{algorithm}[tbp]
\caption{Alignment-conditioned predictive coding of one cache object.}
\label{alg:predictive-kv-codec}
\begin{algorithmic}[1]
\STATE {\bfseries Input.} blocks $\{X_v\}_{v=1}^{m}$, frozen $\{\mu_v,A_v\}$, target $\epsilon$, and $\lambda_2=0.5$, $\lambda_v=0.75$ for $v\ge3$
\STATE Align $Z_v=(X_v-\mathbf{1}\mu_v^\top)A_v$ for $v=1,\ldots,m$
\STATE Set $\eta=\epsilon\|X\|_F/\sqrt n$ as in Equation~\ref{eq:method-relative-error}
\STATE Store $b_c=\frac{1}{mN}\sum_{v=1}^{m}\sum_{t=0}^{N-1}z_{v,t,c}$
\FOR{each channel $c$}
\STATE {\bfseries Block $1$, even $t$.} $\widehat z_{1,t,c}\leftarrow\textsc{Code}(z_{1,t,c},\,b_c)$
\STATE {\bfseries Block $1$, odd $t$.} $\widehat z_{1,t,c}\leftarrow\textsc{Code}\big(z_{1,t,c},\,(\widehat z_{1,t-1,c}+\widehat z_{1,t+1,c})/2\big)$; the last odd site repeats its left neighbor
\FOR{$v=2,\ldots,m$}
\FOR{$t=0,\ldots,N-1$}
\STATE $a_{v,t,c}=\frac{1}{v-1}\sum_{j=1}^{v-1}\widehat z_{j,t,c}$,\quad $p_{v,t,c}=b_c+\lambda_v(a_{v,t,c}-b_c)$
\STATE $\widehat z_{v,t,c}\leftarrow\textsc{Code}(z_{v,t,c},\,p_{v,t,c})$
\ENDFOR
\ENDFOR
\ENDFOR
\STATE Entropy-code symbols with rANS~\citep{duda2013ans}
\STATE {\bfseries Decode.} recover symbols and escapes, repeat the same prediction order, and restore $\widehat X_v=\widehat Z_vA_v^\top+\mathbf{1}\mu_v^\top$
\STATE {\bfseries Procedure} $\textsc{Code}(z,p)$: emit one symbol and return the reconstruction $\widehat z$
\STATE \quad $r=z-p$, $u=\lfloor|r|/\eta\rfloor+1$, $k=\lfloor u/2\rfloor$, $s=\mathrm{sign}(r)$
\STATE \quad \textbf{if} $u<256$ and $|p+2sk\eta-z|\le\eta$: emit $q=128+sk$; \textbf{return} $\widehat z=p+2(q-128)\eta$
\STATE \quad \textbf{else}: emit escape $q=0$ and store $z$ as FP32; \textbf{return} $\widehat z=z$
\end{algorithmic}
\end{algorithm}

Equation~\ref{eq:method-relative-error} sets the residual step from the
requested bound. Varying $\epsilon$ gives different storage sizes and
measured reconstruction errors. Prediction references use FP32, with wider accumulation
intermediates. Final measured errors include inverse transformation and
format conversion.

\subsection{Alignment fitting}
\label{sec:gpa-fitting-details}
\paragraph{Input.}
Calibration matrices $S_v\in\mathbb{R}^{n_c\times d}$ for $v=1,\ldots,m$ in
one group.

\paragraph{Preprocessing.}
Set $\mu_v=\frac1{n_c}S_v^\top\mathbf{1}$. Center and normalize each row
$\bar s_{v,t}=(s_{v,t}-\mu_v^\top)/\max(\|s_{v,t}-\mu_v^\top\|_2,10^{-8})$.
Normalization is used only for fitting.

\paragraph{Alternating updates.}
Start from $A_v=I$. Independently per group, iterate
\begin{equation}
 M=\frac1m\sum_v\bar S_vA_v,\qquad
 \bar S_v^\top M=U_v\Sigma_vV_v^\top,\qquad
 \widetilde A_v=U_vV_v^\top,\quad
 A_v=\widetilde A_v\widetilde A_1^\top.
 \label{eq:method-procrustes-update}
\end{equation}
The last assignment anchors the first block.

\paragraph{Stopping criterion.}
At most 100 iterations with tolerance $10^{-6}$. Cross-products are
precomputed.

\paragraph{Stored parameters.}
Freeze $\{\mu_v,A_v\}$ for compression. Row normalization is not applied
when forming $Z_v$ in Algorithm~\ref{alg:predictive-kv-codec}.

\paragraph{Fitting complexity.}
For $n_c$ samples per block, precomputing pairwise cross-products costs
$O(n_c m^2d^2)$ time and $O(m^2d^2)$ storage. Each GPA iteration forms
$m$ cross-covariance sums and computes $m$ SVDs of size $d\times d$,
costing $O(m^2d^3)$ time in total, including $O(md^3)$ for the SVDs.
With $I$ iterations, fitting therefore costs
$O(n_c m^2d^2+I m^2d^3)$. Cross-products are temporary fitting workspace;
the deployed rotation bank contains only $O(md^2)$ entries.

\subsection{Cache organization and metadata}
\label{sec:frame-implementation}
Objects contain 256 cache positions. A group has shape $d\times m\times256$.
Channel-wise serialization visits even then odd positions of its first block,
followed by subsequent blocks. Dependencies stay within a group and channel.
Table~\ref{tab:cache-layout} specifies the offline RD configuration.

\begin{table}[tbp]
\centering\small\setlength{\tabcolsep}{5pt}
\caption{Cache geometry for Figure~\ref{fig:measured-rd}.}
\label{tab:cache-layout}
\begin{tabular}{llrrl}
\toprule
Model & Compressed cache & $d$ & Streams/object & Blocks/group \\
\midrule
Qwen3.5-9B & Full-attention K/V & 256 & 16 & 4 heads each \\
GLM-5.3-Flash & Post-LN latent & 512 & 3 & 4, 4, 3 layers \\
DeepSeek-V4-Flash & C4 CSA latent & 512 & 6 & $1+5\times4$ layers \\
\bottomrule
\end{tabular}
\end{table}

DeepSeek's zero-based layer groups are $[2]$, $[4,6,8,10]$,
$[12,14,16,18]$, $[20,22,24,26]$, $[28,30,32,34]$, and $[36,38,40,42]$.
The singleton uses the auxiliary mean, identity rotation, and intra-block
prediction. Each remaining group fits GPA independently. All groups use
40,960 auxiliary positions. Each stream stores geometry, the quantization
threshold, channel prediction means, escape counts and values, and an rANS
payload. Object headers identify the shared parameter bank and stream boundaries.
Incomplete objects retain their original values and count toward bpv.
These partitions follow adjacent native streams with leftover remainder,
without evaluation-set tuning (Appendix~\ref{sec:grouping-sensitivity}).

\subsection{Attention compatibility}
\label{sec:appendix-gqa}
Serving measurements restore each cached token with an explicit inverse.
When the attention path permits, the rotation and mean can move into later
linear maps. For a reconstructed row $\hat x=zA^\top+\mu^\top$ and a
projection with weight $W$ and bias $b$,
\begin{equation}
    \begin{aligned}
        (zA^\top+\mu^\top)W+b^\top &= zW'+(b')^\top,\\
        W'&=A^\top W,\qquad b'=b+W^\top\mu.
    \end{aligned}
    \label{eq:system-affine-absorption}
\end{equation}
For grouped-query attention~\citep{ainslie2023gqa}, query head $r$ attends
to KV head $h=h(r)$, and
$\widehat X_h^c=\widetilde X_h^c(A_h^c)^\top+\mathbf1(\mu_h^c)^\top$
for $c\in\{K,V\}$. The maps must be fixed on the attended prefix and on
appended tokens. Post-RoPE queries use
\begin{equation}
 \widetilde q_r=q_rA_h^K,\qquad
 q_r\widehat K_h^\top=\widetilde q_r\widetilde K_h^\top
 +(q_r\mu_h^K)\mathbf1^\top.
 \label{eq:system-gqa-key}
\end{equation}
The mean term cancels in softmax. Rotation follows RoPE. With attention
weights $p_r$ satisfying $p_r\mathbf1=1$,
\begin{equation}
 z_r=p_r\widetilde V_h,\qquad
 u_r=p_r\widehat V_h=z_r(A_h^V)^\top+(\mu_h^V)^\top.
 \label{eq:system-gqa-value}
\end{equation}
A query-dependent gate $\gamma_r$ generally blocks static absorption into
the output projection. Ungated GQA can absorb the inverse and mean. We do
not use this path.

\section{Detailed Experimental Results}
\label{sec:measured-domain-bpv}

\subsection{Metrics and baseline settings}
\label{sec:measured-metrics}
For cache objects $X_i$, reconstructions $\widehat X_i$, and serialized byte
counts $S_i$, the offline RD metrics are
\begin{equation}
 \mathrm{bpv}=\frac{8\sum_iS_i}{\sum_i|X_i|},\qquad
 \mathrm{NRMSE}=\sqrt{\frac{\sum_i\|\widehat X_i-X_i\|_F^2}{\sum_i\|X_i\|_F^2}}.
\end{equation}
They aggregate all five evaluation documents per model. Stored sizes include
metadata and headers but exclude reusable fitted parameters. Full-context
quality reports official task score versus storage rate, with the accounting
described in Appendix~\ref{sec:quality-measurements}.
Figure~\ref{fig:quality-rd-appendix} additionally plots mean per-example
compressed-region NRMSE.

\begin{table}[tbp]
\centering\small
\caption{Offline RD baseline and operating-point settings.}
\label{tab:rd-protocol}
\begin{tabular}{@{}p{0.24\linewidth}p{\dimexpr0.76\linewidth-2\tabcolsep\relax}@{}}
\toprule
Item & Setting \\
\midrule
Input caches & All methods cover the same cache positions and retain positional encoding where applicable, keeping their input representations consistent. \\
FP8 reference & E4M3FN, scale 1. FP16 inputs clamped to $[-448,448]$. \\
KVCodec & Adapted software-codec reference: unrotated token layout, lossless x265, veryfast preset, three B-frames with adaptive selection, and at most 256 positions per GOP. \\
KVTC-style & Adapted transform-coding reference: joint PCA, quantization, and Deflate. DeepSeek uses three-layer groups. Shared PCA parameters are excluded from bpv. \\
SZ3 auto & Official SZ3 auto on the same caches and frozen means as Ours. Rotation + SZ3 applies fitted rotations, and unrotated SZ3 auto does not. Figure~\ref{fig:measured-rd}, Table~\ref{tab:matched-storage}, Table~\ref{tab:rd-detail-sz3}, and Table~\ref{tab:rd-detail-sz3-identity}. \\
Residual-only reference & Component-ablation control (Table~\ref{tab:component-measured}, first row): error-lattice quantization plus rANS, with alignment and prediction disabled. \\
Operating points & Keep fitted parameters fixed and adjust only baseline rate controls on evaluation caches toward Ours' measured error. Figure~\ref{fig:measured-rd} is the primary comparison. Table~\ref{tab:matched-storage} summarizes configurations near the FP8 reference, reporting each method's measured error rather than requiring an exact match. \\
\bottomrule
\end{tabular}
\end{table}

KVTC-style and KVCodec in the figures and tables denote the adapted
implementations above. The comparison uses common target cache positions
and comparable measured reconstruction error, while retaining method-specific
transforms, quantization policies, and codecs. The same-host CPU timings in
Appendix~\ref{sec:cpu-codec-throughput} measure the cost of these codec
implementations. End-to-end serving measurements evaluate AlignPred within
HiCache (Appendix~\ref{sec:appendix-system}).

\subsection{Related-method positioning}
\label{sec:related-positioning}
Table~\ref{tab:related-positioning} classifies nearby methods by primary
setting and their role in the experiments reported here. Persistent KV-cache
storage comparisons use measured reconstruction error and full-cache bpv on
the InfiniteBench caches. LongBench-Pro evaluates how reconstruction-error
operating points affect task score.

\begin{table}[tbp]
\centering\small\setlength{\tabcolsep}{4pt}
\caption{Nearby methods by primary setting and experimental role.}
\label{tab:related-positioning}
\begin{tabular}{@{}p{0.24\linewidth}p{0.28\linewidth}p{0.42\linewidth}@{}}
\toprule
Method & Primary setting & Role here \\
\midrule
CacheGen & Remote KV streaming & Related persistent/remote KV system \\
KVTC & Persistent KV storage & Adapted transform-coding reference \\
KVCodec & Remote KV reuse & Adapted software-codec reference \\
LLM.265 & Video codec as tensor codec & Related codec-style KV compression \\
SZ3 auto & Error-bounded array compression & Generic error-controlled baseline \\
KIVI, TurboQuant & GPU-resident KV quantization & Task quality at different storage sizes \\
KVQuant, RotateKV & GPU low-bit KV & Related GPU-resident quantization methods \\
Palu, MatryoshkaKV, xKV & Rank / feature reduction & Related alignment-aware compression \\
\bottomrule
\end{tabular}
\end{table}

\paragraph{Local transforms versus KVTC.}
KVTC~\citep{staniszewski2026kvtc} exploits cross-block redundancy through
joint PCA; AlignPred uses local alignment followed by prediction.
Table~\ref{tab:local-joint-complexity} compares their transform costs.
Our transform width remains $d$ as group size changes, allowing groups to
follow native head or layer boundaries, including singleton blocks.

\begin{table}[tbp]
\centering\small\setlength{\tabcolsep}{4pt}
\caption{Local transforms versus full-width joint PCA for $m$ blocks of
width $d$ and $N$ cache positions.}
\label{tab:local-joint-complexity}
\begin{tabular}{@{}p{0.30\linewidth}p{0.31\linewidth}p{0.33\linewidth}@{}}
\toprule
 & AlignPred & Full-width joint PCA \\
\midrule
Transform matrices & $m$ matrices, $d\times d$ & One matrix, $md\times md$ \\
Parameter storage & $O(md^2)$ & $O(m^2d^2)$ \\
Transform work & $O(Nmd^2)$ & $O(Nm^2d^2)$ \\
Decomposition size & $d\times d$ & $md\times md$ \\
Decomposition work & $O(md^3)$ per GPA iteration & $O(m^3d^3)$ \\
\bottomrule
\end{tabular}
\par\smallskip
\begin{minipage}{\linewidth}
\footnotesize
GPA's total fitting cost, including cross-products and iterations, is given
in Section~\ref{sec:gpa-fitting-details}.
KVTC's randomized rank-$r$ PCA uses $O(mdr)$ basis storage and
$O(Nmdr)$ transform work. Relative to our local maps, the matrix-size
ratio is $m$ for the full basis and $r/d$ for the truncated basis.
\end{minipage}
\end{table}

\begin{figure}[H]
    \centering
    \includegraphics[width=\linewidth]{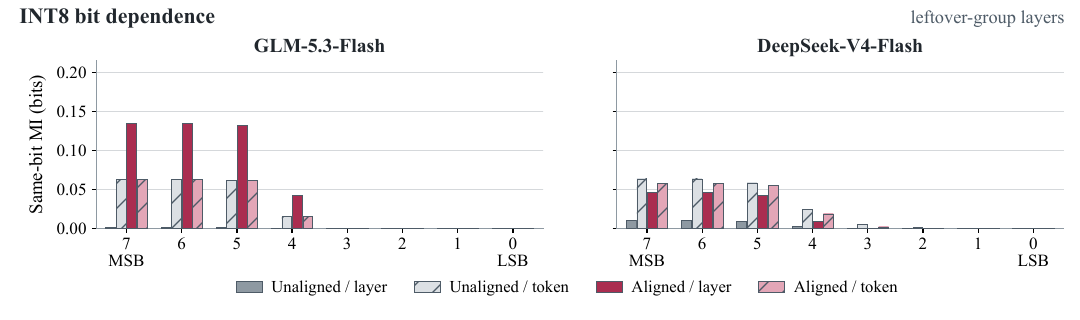}
    \caption{Same INT8 same-bit MI diagnostic as Figure~\ref{fig:bit-dependence}
    on GLM MLA and DeepSeek CSA leftover-group layers.}
    \label{fig:bit-dependence-latent}
\end{figure}

\paragraph{Motivation measurement.}
\label{sec:motivation-experiment-details}
Figure~\ref{fig:bit-dependence} fits centered GPA on a 16,384-token
InfiniteBench document and evaluates a separate 32,768-token document with
Qwen3.5-9B. Both cache representations use the same means and error target 0.03.
Same-bit mutual information is accumulated over tokens per channel and
neighboring head pair (or neighboring token pair), then averaged. Keys and
Values are reported separately.
The same protocol is repeated on GLM-5.3-Flash MLA leftovers $4/4/3$ and
on DeepSeek-V4-Flash CSA leftovers $1{+}5\times4$, where neighboring
leftover-group layers play the role of Qwen heads.
Figure~\ref{fig:bit-dependence-latent} uses the same INT8 codes, error
target, and y-scale.
After alignment, GLM high-order neighboring-layer MI exceeds
neighboring-token MI, as in Figure~\ref{fig:bit-dependence}.
DeepSeek neighboring-layer MI also rises.

\FloatBarrier

\subsection{Complete storage and reconstruction error measurements}
Figure~\ref{fig:measured-rd} compares storage size at different reconstruction errors.
Table~\ref{tab:matched-storage} summarizes representative operating points
near the FP8 reconstruction-error reference from
Section~\ref{sec:numerical-rd}, following Table~\ref{tab:rd-protocol}.
Methods have different rate controls and attainable errors, so the table
reports both stored bpv and measured NRMSE for each configuration.
\begin{table}[H]
\centering
\small
\caption{Representative operating points near the FP8 reconstruction-error
reference, reported as bpv (NRMSE, \%).
Figure~\ref{fig:measured-rd} is the primary rate--distortion comparison.
SZ3 uses auto mode on the same locally aligned caches.}
\label{tab:matched-storage}
\begin{tabular}{lrrrr}
\toprule
Model & Ours & KVCodec & KVTC-style & Rotation + SZ3 \\
\midrule
Qwen3.5-9B & \textbf{4.347} (2.67) & 5.642 (2.51) & 4.931 (2.58) & 4.775 (2.60) \\
GLM-5.3-Flash & \textbf{5.025} (2.66) & 6.035 (2.66) & 5.909 (3.07) & 5.395 (2.59) \\
DeepSeek-V4-Flash & \textbf{5.049} (2.66) & 5.547 (2.66) & 5.974 (3.03) & 5.404 (2.60) \\
\bottomrule
\end{tabular}
\end{table}

Tables~\ref{tab:rd-detail-qwen}--\ref{tab:rd-detail-deepseek} give measured
Figure~\ref{fig:measured-rd} points for Ours, KVCodec, and KVTC-style.
Table~\ref{tab:rd-detail-sz3} gives Rotation + SZ3 auto, and
Table~\ref{tab:rd-detail-sz3-identity} gives unrotated SZ3 auto, from the
same figure.
$m$ denotes the calibration-error multiplier, $b$ the KVTC-style payload bits before
entropy, and Ref. an additional rate-adjusted configuration near the FP8
reference. Error is pooled NRMSE (\%).
Methods are independently sorted by bpv. Rows do not imply matched errors.
\begin{table}[tbp]
\centering
\caption{Qwen3.5-9B: numerical rate--distortion results for Figure~\ref{fig:measured-rd}. FP8 NRMSE: 2.65\%.}
\label{tab:rd-detail-qwen}
\small\setlength{\tabcolsep}{5pt}\renewcommand{\arraystretch}{0.95}
\begin{tabular}{rrr@{\hspace{14pt}}rrr@{\hspace{14pt}}rrr}
\toprule
\multicolumn{3}{c}{Ours} & \multicolumn{3}{c}{KVCodec} & \multicolumn{3}{c}{KVTC-style} \\
\cmidrule(lr){1-3}\cmidrule(lr){4-6}\cmidrule(lr){7-9}
$m$ & bpv & Error (\%) & $m$ & bpv & Error (\%) & $b$ & bpv & Error (\%) \\
\midrule
4 & 2.398 & 10.61 & 8 & 2.224 & 21.05 & 3.5 & 2.632 & 16.52 \\
3 & 2.787 & 7.98 & 6.452 & 2.499 & 17.13 & 4 & 3.186 & 9.58 \\
2 & 3.352 & 5.33 & 6 & 2.593 & 15.96 & 4.5 & 3.610 & 5.91 \\
1.5 & 3.761 & 4.00 & 4 & 3.137 & 10.69 & 6 & 4.931 & 2.58 \\
1.25 & 4.024 & 3.33 & 3 & 3.536 & 8.04 & 7 & 5.760 & 2.13 \\
1 & 4.347 & 2.67 & 2 & 4.112 & 5.41 & -- & -- & -- \\
0.75 & 4.767 & 2.00 & 1.5 & 4.538 & 4.13 & -- & -- & -- \\
0.5 & 5.362 & 1.33 & 1.25 & 4.811 & 3.55 & -- & -- & -- \\
-- & -- & -- & 1 & 5.143 & 3.03 & -- & -- & -- \\
-- & -- & -- & 0.75 & 5.542 & 2.59 & -- & -- & -- \\
-- & -- & -- & Ref. & 5.642 & 2.51 & -- & -- & -- \\
-- & -- & -- & 0.5 & 5.919 & 2.32 & -- & -- & -- \\
\bottomrule
\end{tabular}
\end{table}

\begin{table}[tbp]
\centering
\caption{GLM-5.3-Flash: numerical rate--distortion results for Figure~\ref{fig:measured-rd}. FP8 NRMSE: 2.66\%.}
\label{tab:rd-detail-glm}
\small\setlength{\tabcolsep}{5pt}\renewcommand{\arraystretch}{0.95}
\begin{tabular}{rrr@{\hspace{14pt}}rrr@{\hspace{14pt}}rrr}
\toprule
\multicolumn{3}{c}{Ours} & \multicolumn{3}{c}{KVCodec} & \multicolumn{3}{c}{KVTC-style} \\
\cmidrule(lr){1-3}\cmidrule(lr){4-6}\cmidrule(lr){7-9}
$m$ & bpv & Error (\%) & $m$ & bpv & Error (\%) & $b$ & bpv & Error (\%) \\
\midrule
6 & 2.479 & 15.96 & 12 & 2.199 & 32.03 & 3.5 & 2.775 & 24.64 \\
4 & 3.037 & 10.65 & 9.537 & 2.494 & 25.46 & 4 & 3.301 & 16.70 \\
3 & 3.443 & 7.99 & 8 & 2.730 & 21.37 & 4.5 & 3.804 & 10.75 \\
2 & 4.022 & 5.33 & 6 & 3.136 & 16.02 & 6 & 5.087 & 3.96 \\
1.5 & 4.437 & 4.00 & 4 & 3.733 & 10.69 & 7 & 5.909 & 3.07 \\
1.25 & 4.701 & 3.33 & 3 & 4.176 & 8.01 & -- & -- & -- \\
1 & 5.025 & 2.66 & 2 & 4.828 & 5.34 & -- & -- & -- \\
0.75 & 5.445 & 2.00 & 1.5 & 5.315 & 4.01 & -- & -- & -- \\
0.5 & 6.040 & 1.33 & 1.25 & 5.633 & 3.34 & -- & -- & -- \\
-- & -- & -- & 1 & 6.030 & 2.67 & -- & -- & -- \\
-- & -- & -- & Ref. & 6.035 & 2.66 & -- & -- & -- \\
-- & -- & -- & 0.75 & 6.556 & 2.00 & -- & -- & -- \\
-- & -- & -- & 0.5 & 7.309 & 1.34 & -- & -- & -- \\
\bottomrule
\end{tabular}
\end{table}

\begin{table}[tbp]
\centering
\caption{DeepSeek-V4-Flash: numerical rate--distortion results for Figure~\ref{fig:measured-rd}. FP8 NRMSE: 2.67\%.}
\label{tab:rd-detail-deepseek}
\small\setlength{\tabcolsep}{5pt}\renewcommand{\arraystretch}{0.95}
\begin{tabular}{rrr@{\hspace{14pt}}rrr@{\hspace{14pt}}rrr}
\toprule
\multicolumn{3}{c}{Ours} & \multicolumn{3}{c}{KVCodec} & \multicolumn{3}{c}{KVTC-style} \\
\cmidrule(lr){1-3}\cmidrule(lr){4-6}\cmidrule(lr){7-9}
$m$ & bpv & Error (\%) & $m$ & bpv & Error (\%) & $b$ & bpv & Error (\%) \\
\midrule
8 & 2.128 & 21.00 & 8 & 2.412 & 20.96 & 3.5 & 2.864 & 27.21 \\
6.366 & 2.429 & 16.81 & 6 & 2.794 & 15.76 & 4 & 3.413 & 18.67 \\
6 & 2.509 & 15.86 & 4 & 3.361 & 10.53 & 4.5 & 3.886 & 11.24 \\
4 & 3.065 & 10.62 & 3 & 3.783 & 7.92 & 6 & 5.111 & 4.22 \\
3 & 3.468 & 7.97 & 2 & 4.406 & 5.29 & 7 & 5.974 & 3.03 \\
2 & 4.046 & 5.32 & 1.5 & 4.868 & 3.97 & -- & -- & -- \\
1.5 & 4.460 & 3.99 & 1.25 & 5.166 & 3.32 & -- & -- & -- \\
1.25 & 4.724 & 3.33 & 1 & 5.517 & 2.70 & -- & -- & -- \\
1 & 5.049 & 2.66 & Ref. & 5.547 & 2.66 & -- & -- & -- \\
0.75 & 5.469 & 2.00 & 0.75 & 5.919 & 2.21 & -- & -- & -- \\
0.5 & 6.065 & 1.33 & 0.5 & 6.277 & 1.95 & -- & -- & -- \\
\bottomrule
\end{tabular}
\end{table}

\begin{table}[tbp]
\centering
\caption{Rotation+SZ3 auto points for Figure~\ref{fig:measured-rd}, on the
same locally aligned caches and frozen means/rotations as Ours.
$m=1$ is the Table~\ref{tab:matched-storage} column.}
\label{tab:rd-detail-sz3}
\small\setlength{\tabcolsep}{5pt}\renewcommand{\arraystretch}{0.95}
\begin{tabular}{rrr@{\hspace{14pt}}rrr@{\hspace{14pt}}rrr}
\toprule
\multicolumn{3}{c}{Qwen3.5-9B} & \multicolumn{3}{c}{GLM-5.3-Flash} & \multicolumn{3}{c}{DeepSeek-V4-Flash} \\
\cmidrule(lr){1-3}\cmidrule(lr){4-6}\cmidrule(lr){7-9}
$m$ & bpv & Error (\%) & $m$ & bpv & Error (\%) & $m$ & bpv & Error (\%) \\
\midrule
3 & 3.193 & 7.79 & 8 & 2.427 & 20.71 & 8 & 2.439 & 20.65 \\
2 & 3.769 & 5.20 & 4 & 3.389 & 10.36 & 4 & 3.400 & 10.37 \\
1.5 & 4.187 & 3.90 & 2 & 4.378 & 5.18 & 2 & 4.386 & 5.19 \\
1.25 & 4.450 & 3.25 & 1 & 5.395 & 2.59 & 1 & 5.404 & 2.60 \\
1 & 4.775 & 2.60 & 0.5 & 6.402 & 1.30 & 0.5 & 6.406 & 1.30 \\
0.75 & 5.205 & 1.95 & -- & -- & -- & -- & -- & -- \\
0.5 & 5.798 & 1.30 & -- & -- & -- & -- & -- & -- \\
\bottomrule
\end{tabular}
\end{table}

\begin{table}[tbp]
\centering
\caption{Unrotated SZ3 auto points for Figure~\ref{fig:measured-rd}, on the
same caches and frozen means as Ours, without fitted rotations.
$m=1$ matches Table~\ref{tab:rd-detail-sz3}.}
\label{tab:rd-detail-sz3-identity}
\small\setlength{\tabcolsep}{5pt}\renewcommand{\arraystretch}{0.95}
\begin{tabular}{rrr@{\hspace{14pt}}rrr@{\hspace{14pt}}rrr}
\toprule
\multicolumn{3}{c}{Qwen3.5-9B} & \multicolumn{3}{c}{GLM-5.3-Flash} & \multicolumn{3}{c}{DeepSeek-V4-Flash} \\
\cmidrule(lr){1-3}\cmidrule(lr){4-6}\cmidrule(lr){7-9}
$m$ & bpv & Error (\%) & $m$ & bpv & Error (\%) & $m$ & bpv & Error (\%) \\
\midrule
3 & 3.481 & 7.80 & 8 & 2.668 & 20.72 & 8 & 2.544 & 20.47 \\
2 & 4.058 & 5.20 & 4 & 3.632 & 10.36 & 4 & 3.511 & 10.30 \\
1.5 & 4.477 & 3.90 & 2 & 4.628 & 5.18 & 2 & 4.504 & 5.17 \\
1.25 & 4.742 & 3.25 & 1 & 5.650 & 2.59 & 1 & 5.524 & 2.59 \\
1 & 5.069 & 2.60 & 0.5 & 6.649 & 1.30 & 0.5 & 6.535 & 1.30 \\
0.75 & 5.493 & 1.95 & -- & -- & -- & -- & -- & -- \\
0.5 & 6.082 & 1.30 & -- & -- & -- & -- & -- & -- \\
\bottomrule
\end{tabular}
\end{table}

\subsection{Complete downstream-quality measurements}
\label{sec:quality-measurements}
\begin{figure}[tbp]
\centering
\includegraphics[width=\linewidth]{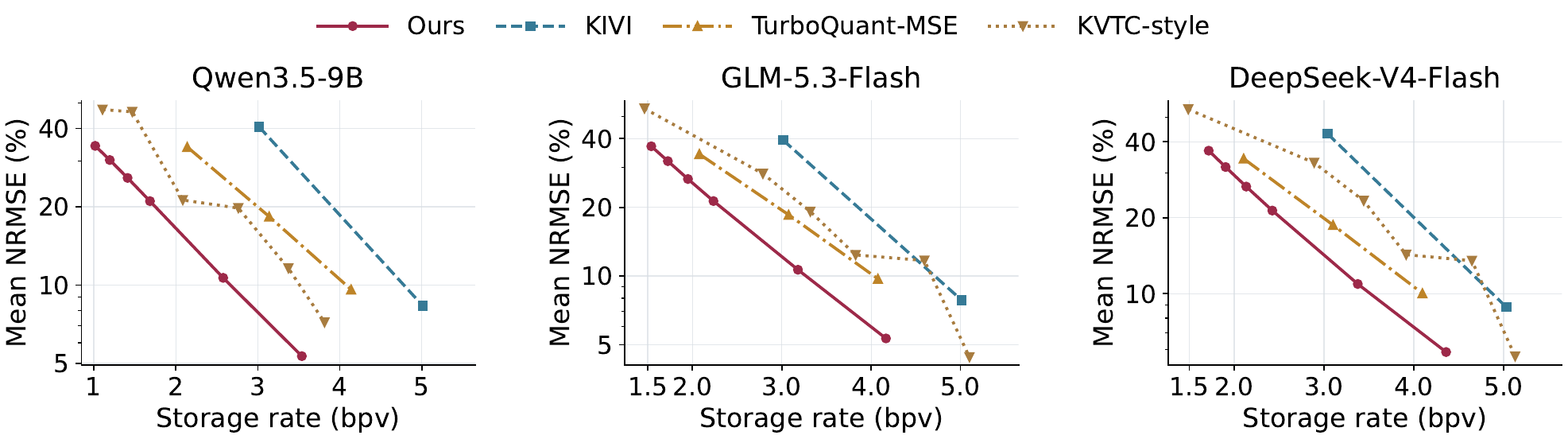}
\caption{Reconstruction-error companion to Figure~\ref{fig:task-quality}:
the same LongBench-Pro compression settings, plotted as mean
compressed-region NRMSE (\%) versus storage rate (bpv, including tails and
metadata). Figure~\ref{fig:task-quality} uses task score on the vertical
axis. This figure uses numerical fidelity. Error axes are logarithmic.}
\label{fig:quality-rd-appendix}
\end{figure}

\begin{figure}[tbp]
\centering
\includegraphics[width=\linewidth,height=0.88\textheight,keepaspectratio]{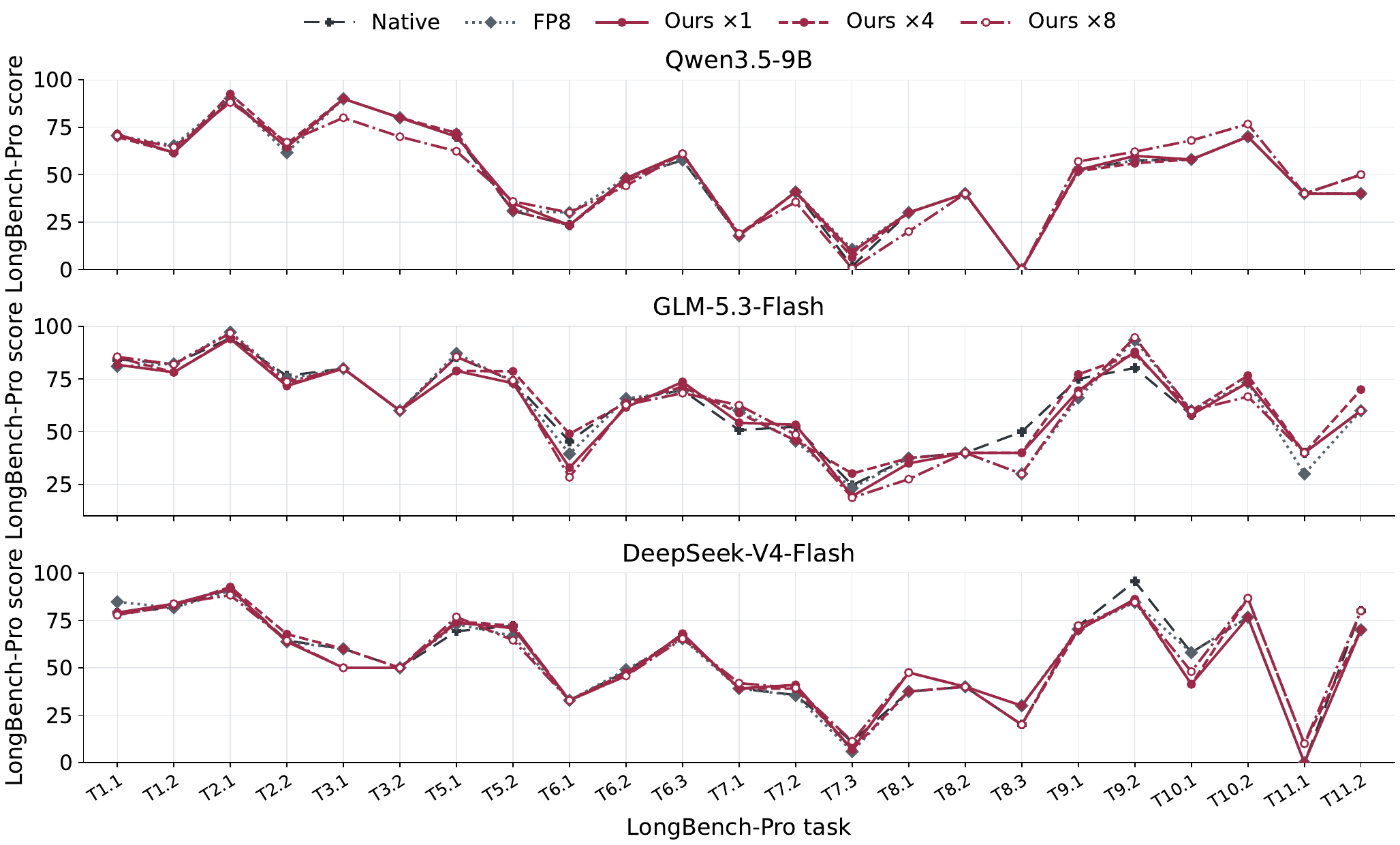}
\caption{Per-task official LongBench-Pro scores for Native, FP8, and Ours
at error multipliers $m=1,4,8$, i.e.\ the AlignPred settings highlighted
in Figure~\ref{fig:task-quality}. Each panel is one model.}
\label{fig:quality-per-task}
\end{figure}

Figure~\ref{fig:quality-rd-appendix} plots compressed-region reconstruction
error versus storage rate for the same LongBench-Pro settings as
Section~\ref{sec:task-quality}.
Figure~\ref{fig:quality-per-task} reports per-task scores.

Task scores and mean reconstruction errors use all examples in the
LongBench-Pro 128K setting except the summarization tasks.

\subsection{Calibration strategies}
\label{sec:calibration-details}
Auxiliary fitting uses 40,960 positions from five independent documents.
Cache-specific fitting uses 8,192 positions per target document. Both use
the GPA routine above and retain the auxiliary error reference at all rates.
Table~\ref{tab:calibration-strategies} pools the five InfiniteBench evaluation
documents (same caches as Table~\ref{tab:matched-storage}).
Figure~\ref{fig:calibration-strategies} instead uses one English cache per
model, matching Table~\ref{tab:component-measured}.
The five-category caches have 98,304 input tokens, except GLM/DeepSeek
mathematics at 90,000. DeepSeek's C4 pool has 24,576 or 22,500 cache positions.
Bytes and squared errors are pooled across documents. Equal multipliers do
not enforce exactly equal errors (Qwen at $m=8$: 20.58\% versus 20.20\%).
\begin{table}[tbp]
    \caption{Five-category calibration storage (bpv). ``Specific total''
    includes one dedicated parameter bank per document.}
    \label{tab:calibration-strategies}
    \centering
    \small
    \begin{tabular}{llrrr}
        \toprule
        Model & Multiplier & Auxiliary & Specific & Specific total \\
        \midrule
        Qwen3.5-9B & 1 & 4.347 & 4.132 & 4.216 \\
         & 4 & 2.398 & 2.210 & 2.294 \\
         & 8 & 1.525 & 1.369 & 1.453 \\
        GLM-5.3-Flash & 1 & 5.025 & 4.901 & 5.071 \\
         & 4 & 3.037 & 2.920 & 3.090 \\
         & 8 & 2.100 & 1.995 & 2.165 \\
        DeepSeek-V4-Flash & 1 & 5.049 & 4.944 & 5.623 \\
         & 4 & 3.065 & 2.966 & 3.645 \\
         & 8 & 2.128 & 2.039 & 2.718 \\
        \bottomrule
    \end{tabular}
\end{table}

FP32 matrices and means, including the anchored identity and serialization
headers, contribute approximately $32(d+1)/T$ bpv for channel dimension $d$
and $T$ cache positions. Fitting is performed once per bank. Times include
GPU transfers, centering, normalization, cross-products and GPA, but exclude
source reading and sampling. Auxiliary fitting is offline. Coding uses
32 CPU threads. Parameter overhead is charged per document only for the
cache-specific bank. Mean per-document fitting times are 14.82, 7.55,
and 14.21 seconds for Qwen, GLM, and DeepSeek, respectively.

\subsection{Blending-weight sensitivity}
\label{sec:lambda-ablation}
The default blending weights, $\lambda_2=0.5$ and $\lambda_{v\ge 3}=0.75$,
were selected on independent development data and fixed before the
InfiniteBench evaluation. All main-paper results use these fixed weights
without tuning on the five held-out evaluation documents. To assess
sensitivity to this choice, we perform a post-hoc sweep over
$\{0,0.25,0.5,0.75,1\}^2$ on Qwen3.5-9B, keeping GPA, intra-block
interpolation, the residual lattice, and $\eta$ fixed. Bytes and squared
errors are pooled as in the calibration study. The sweep uses the five
InfiniteBench held-out documents and the frozen auxiliary bank of
Section~\ref{sec:experiments}. Table~\ref{tab:lambda-ablation-x1} reports
stored bpv at multiplier $m=1$, where measured NRMSE is $2.667\%$ in every
cell.

\begin{table}[tbp]
\centering\small
\caption{Post-hoc blending-weight sensitivity on Qwen at $m=1$ (bpv
$\downarrow$). Bold marks the pre-specified default, fixed on independent
development data before evaluation. Results are pooled over five held-out
InfiniteBench documents at matched reconstruction error.}
\label{tab:lambda-ablation-x1}
\setlength{\tabcolsep}{4pt}
\begin{tabular}{crrrrr}
\toprule
& \multicolumn{5}{c}{$\lambda_{v\ge 3}$} \\
\cmidrule(lr){2-6}
$\lambda_2$ & $0$ & $0.25$ & $0.5$ & $0.75$ & $1$ \\
\midrule
$0$ & 4.644 & 4.547 & 4.473 & 4.437 & 4.452 \\
$0.25$ & 4.599 & 4.497 & 4.419 & 4.381 & 4.397 \\
$0.5$ & 4.572 & 4.467 & 4.387 & \textbf{4.347} & 4.363 \\
$0.75$ & 4.570 & 4.464 & 4.383 & 4.342 & 4.358 \\
$1$ & 4.594 & 4.491 & 4.411 & 4.371 & 4.386 \\
\bottomrule
\end{tabular}

\end{table}

The pre-specified default is robust to moderate changes in the blending
weights. At $m=1$, the best tested setting,
$\lambda_2=\lambda_{v\ge 3}=0.75$, improves storage by only $0.005$~bpv
over the default ($4.342$ vs.\ $4.347$~bpv). At $m=4$ and $m=8$, the
default attains the lowest rate among the tested settings
(Table~\ref{tab:lambda-ablation-rates}). Setting both weights to $0$ removes
cross-block prediction and costs $0.22$--$0.30$~bpv, while setting both to
$1$ removes shrinkage and costs $0.038$--$0.070$~bpv. These results show
that the benefit of cross-block prediction does not depend on precise tuning
of the blending weights.

\begin{table}[tbp]
\centering\small
\caption{Selected blending-weight settings at three error multipliers (bpv
$\downarrow$). Bold marks the pre-specified default. No weights are tuned
on these evaluation documents.}
\label{tab:lambda-ablation-rates}
\setlength{\tabcolsep}{5pt}
\begin{tabular}{ccrrr}
\toprule
$\lambda_2$ & $\lambda_{v\ge 3}$ & $m=1$ & $m=4$ & $m=8$ \\
\midrule
$0$ & $0$ & 4.644 & 2.665 & 1.740 \\
$0.5$ & $0.75$ & \textbf{4.347} & \textbf{2.398} & \textbf{1.525} \\
$0.75$ & $0.75$ & 4.342 & 2.399 & 1.533 \\
$1$ & $1$ & 4.386 & 2.451 & 1.595 \\
\bottomrule
\end{tabular}

\end{table}

\subsection{Group construction}
\label{sec:grouping-sensitivity}
We intentionally use simple groups inherited from adjacent native cache
streams, without optimizing group partitions on the evaluation set.
The partitions in Table~\ref{tab:cache-layout} follow this rule.
Qwen uses the four GQA KV heads of one layer, with Keys and Values separate.
GLM leftover-partitions 11 MLA layers as $4/4/3$.
DeepSeek leftover-partitions 21 CSA layers as a first-layer singleton and
five groups of four.

Table~\ref{tab:grouping-sensitivity} evaluates alternative groupings post
hoc. Larger groups can provide slightly lower rates, but the differences are
small and do not affect the main conclusions.

\begin{table}[tbp]
\centering\small
\caption{Held-out grouping sensitivity at $m=1$ (bpv $\downarrow$).
Bold is the paper default, and $\star$ is the evaluation minimum. Pooled over
five InfiniteBench documents at the frozen FP8 error.}
\label{tab:grouping-sensitivity}
\setlength{\tabcolsep}{5pt}
\begin{tabular}{llrr}
\toprule
Model & Partition & Views & bpv \\
\midrule
Qwen3.5-9B & 4 heads / layer & 4 & \textbf{4.347} \\
 & 2 adjacent layers & 8 & $4.304^\star$ \\
 & 4 adjacent layers & 16 & 4.360 \\
 & 8 layers, K or V & 32 & 4.529 \\
GLM-5.3-Flash & $4/4/3$ layers & 4 & \textbf{5.025} \\
 & $5/6$ layers & 6 & 5.031 \\
 & $8/3$ layers & 8 & 5.054 \\
 & 11 layers & 11 & $5.004^\star$ \\
DeepSeek-V4-Flash & $1{+}5\times4$ layers & 4 & \textbf{5.049} \\
 & $7{+}7{+}7$ layers & 7 & 5.033 \\
 & $10{+}11$ layers & 11 & 5.031 \\
 & 21 layers & 21 & $4.996^\star$ \\
\bottomrule
\end{tabular}

\end{table}

\section{System and Timing Protocols}
\label{sec:appendix-system}

\subsection{HiCache restoration}
All three backends reconstruct aligned values with persistent CPU workers
(64 for Qwen, 32 for GLM and DeepSeek), then apply inverse transforms on the GPU before writing the native cache.
Qwen restores separate Keys and Values. GLM and DeepSeek restore shared latent
caches once on rank 0 and broadcast them to all eight ranks. Uncompressed
pools retain their native restoration paths. These implementations preserve
the native host-pool allocation and use additional staging memory. The
reported storage reduction does not imply lower allocated host memory.

\begin{table}[tbp]
\centering\small
\caption{Serving restoration paths.}
\label{tab:serving-implementation}
\begin{tabular}{lp{0.72\linewidth}}
\toprule
Model & Compressed path \\
\midrule
Qwen & Four 64-token pages per object, separate K/V streams, and 32 objects per decode batch. Aligned host staging feeds layerwise GPU inverse transforms and FP16 page scatter. \\
GLM & Post-LN MLA latent, grouped 4/4/3 layers. Rank-0 decoding with 32-object batches, two host/GPU slots, GPU inverse transforms, and ordered BF16 broadcasts. \\
DeepSeek & Pre-FP8 C4 CSA values, grouped $1+5\times4$ layers. Rank-0 decoding with eight-object batches, two host/GPU slots, inverse transforms, broadcasts, and native FP8 writeback. \\
Common & In-memory indices, object read-ahead, page mappings, and event-guarded buffer reuse. Raw tails and native sidecar paths are retained. \\
\bottomrule
\end{tabular}
\end{table}
For shared latents, CPU reconstruction and GPU restoration are pipelined
across groups. Overlap between restoration collectives and model collectives
is not assumed. Startup checks compare the optimized CUDA/TP8 paths with
serial references and test page permutations, untouched slots, repeated
requests, buffer lifetimes, and failure propagation.

\subsection{Serving measurement protocol}
The experiment uses an AMD EPYC 9K84 host with NVIDIA H20 GPUs,
SGLang 0.5.19.dev20260914, and PyTorch 2.13.0+cu130.
Table~\ref{tab:serving-protocol} defines the 36-point write-then-rehit
protocol.
Figure~\ref{fig:serving-storage-ttft} uses the $B=1$ L3-hit restore TTFT
from Table~\ref{tab:online-roundtrip-complete}.
Figure~\ref{fig:serving-write-ttft} uses the matching first-request write
TTFT with no prior L3 hit.
Figure~\ref{fig:serving-batch-ttft} and Table~\ref{tab:batch-rehit-ttft}
report all batch sizes.

Restarting the service clears the application-level L1/L2 caches.
We additionally repeated restore measurements after explicitly clearing
the operating-system file cache and observed similar TTFT, indicating
that the reported latency behavior does not depend on warm file-cache reads.

\begin{table}[tbp]
\centering\small
\caption{Serving experiment matrix and controls.}
\label{tab:serving-protocol}
\begin{tabular}{lp{0.72\linewidth}}
\toprule
Item & Setting \\
\midrule
Models & Qwen3.5-9B: FP16, TP1. GLM-5.3-Flash: BF16, TP8. DeepSeek-V4-Flash: BF16 compute, native mixed FP8 cache, TP8. \\
Matrix & Three models $\times$ 32,768/65,536/131,072-token prefixes $\times$ $B=1/2/4/8$ concurrent requests $\times$ native HiCache/Ours $1\times$. \\
Input/output & $B$ distinct InfiniteBench book documents per model, the first $B$ of one fixed prompt bank, identical across arms and lengths. Each prefix is followed by 64 appended IDs. Greedy generation of exactly 256 tokens, ignoring EOS. \\
Phases & Each arm writes its own L3 files online during the first requests' chunked prefill. The writer drains and stops. A fresh reader then serves the same $B$ requests concurrently from L3. \\
Timing & TTFT: request submission to first token, averaged over the $B$ requests. Persistence completion: first submission to the L3 write of the batch's last target-KV page. Reading, decoding, H2D, inverse transforms, communication, and native writeback are included. \\
Graph/warmup & Prefill graphs disabled. Full decode graphs for batch sizes $1$--$B$. Unrelated full-chunk and 64-token warmups. For $B>1$, readers also run $k$-request EXTEND batches ($k=2,\dots,B$) over an unrelated cached prefix. Startup and warmup excluded. \\
Scheduling & Context length $n+8192$, at most $B$ running requests and $B(n+8192)$ total tokens, capped by GPU memory. Readers of both arms emit a finished prefill's first token before launching the next consecutive prefill (\texttt{SGLANG\_DISABLE\_CONSECUTIVE\_PREFILL\_OVERLAP=1}). \\
Isolation & Every rehit request has a complete L3 prefix hit and zero L1/L2 hits, verified per request. Readers do not write L3. File inventories are unchanged after measurement. \\
Storage & Logical target KV file bytes, including payload, necessary metadata, indices and tails. Other pools and reusable calibration parameters are excluded. Sizes refer to file contents rather than filesystem allocation. \\
\bottomrule
\end{tabular}
\end{table}
\subsection{Online write protocol}
\label{sec:appendix-online-write}
\begin{figure}[tbp]
\centering
\includegraphics[width=\linewidth]{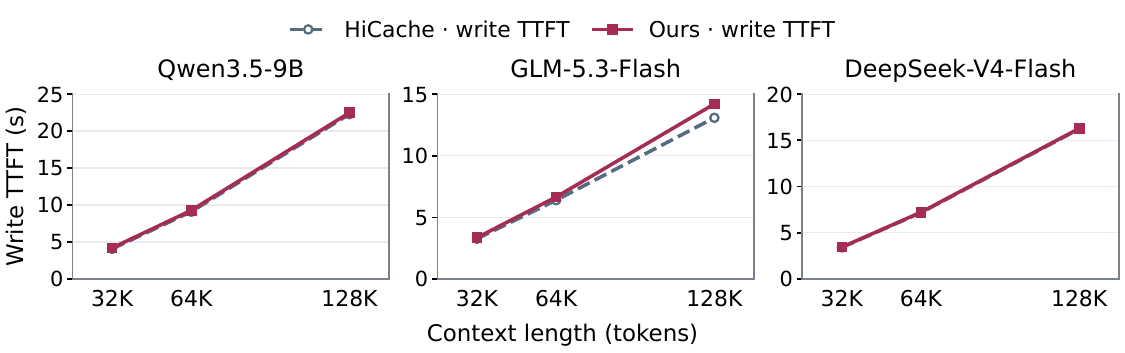}
\caption{First-request write TTFT with no prior L3 hit, while online writes
populate L3. Batch size 1. Axis ranges
vary by model. Storage and L3-hit restore TTFT are in
Figure~\ref{fig:serving-storage-ttft}.}
\label{fig:serving-write-ttft}
\end{figure}

The first requests enable L3 writes:
Qwen uses 8K prefill chunks and GLM and DeepSeek use 2K chunks, with 32 CPU
encode workers. After the write requests, the service is restarted with empty
L1/L2 and restores from those files. Table~\ref{tab:online-roundtrip-64k} reports
the 64K $B=1$ results used in Section~\ref{sec:online-write}.
Table~\ref{tab:online-roundtrip-complete} reports the full 32/64/128K matrix
at $B=1$ and is the source of Figures~\ref{fig:serving-storage-ttft}
and~\ref{fig:serving-write-ttft}.
The input prefixes, 64 appended IDs, 256 generated tokens, H20 host, and
target-KV byte definition match Appendix~\ref{sec:appendix-system}.
DeepSeek's fused store emits FP8 C4, so BF16 C4 values are captured on the
inference stream immediately before each native FP8 store. Qwen and GLM
already expose cached BF16/FP16 at that boundary.

\paragraph{Overlap with chunked prefill.}
\label{sec:appendix-online-overlap}
CPU encoding of a finished chunk runs while later prefill chunks execute.
At 64K, every model starts encoding a finished chunk before a later chunk is
submitted and before the first generated token.
Table~\ref{tab:online-write-overlap} reports 64K overlap on the online writer,
from the $B=1$ write requests in Table~\ref{tab:online-roundtrip-complete}.
CPU wall is the sum of encode-job durations, not critical-path delay:
$8.690$/$2.130$/$2.803$\,s on Qwen/GLM/DeepSeek, while write TTFT rises by
$0.180$/$0.263$/$0.044$\,s versus native HiCache. Qwen persists $87.5$\% of
the prefix before the first token, matching the last $8$K chunk, whose encode
continues into decode. GLM and DeepSeek persist the entire prefix before the
first token.

\begin{table}[tbp]
\centering\small
\setlength{\tabcolsep}{4pt}
\caption{CPU compression overlap at 64K, $B=1$.}
\label{tab:online-write-overlap}
\begin{tabular}{lrrr}
\toprule
Model & CPU wall (s) & $\Delta$ TTFT (s) & Persist before TTFT \\
\midrule
Qwen & 8.690 & 0.180 & 87.5\% \\
GLM & 2.130 & 0.263 & 100.0\% \\
DeepSeek & 2.803 & 0.044 & 100.0\% \\
\bottomrule
\end{tabular}

\end{table}

\begin{table}[tbp]
\centering\small
\caption{64K online write then L3 rehit, $B=1$.}
\label{tab:online-roundtrip-64k}
\begin{tabular}{lrrrrrr}
\toprule
Model & \multicolumn{2}{c}{Write TTFT (s)} & \multicolumn{2}{c}{Rehit TTFT (ms)} & \multicolumn{2}{c}{Stored KV (MiB)} \\
\cmidrule(lr){2-3}\cmidrule(lr){4-5}\cmidrule(lr){6-7}
 & Native & Ours & Native & Ours & Native & Ours \\
\midrule
Qwen & 9.105 & 9.286 & 843.3 & 469.5 & 2048.0 & 577.1 \\
GLM & 6.390 & 6.653 & 628.0 & 502.1 & 704.0 & 227.7 \\
DeepSeek & 7.176 & 7.221 & 316.4 & 302.0 & 192.0 & 110.8 \\
\bottomrule
\end{tabular}

\end{table}

\begin{table}[tbp]
\centering\small
\setlength{\tabcolsep}{2.5pt}
\caption{Online write then L3 rehit, $B=1$. Storage covers the
target KV portion only. Figure~\ref{fig:serving-storage-ttft} plots L3-hit
restore TTFT. Figure~\ref{fig:serving-write-ttft} plots first-request write
TTFT with no L3 hit. Persistence completion is in the $B=1$ column of
Table~\ref{tab:batch-rehit-ttft}(b).}
\label{tab:online-roundtrip-complete}
\begin{tabular}{llrrrrrr}
\toprule
Model & Ctx & \multicolumn{2}{c}{Write TTFT (ms)} & \multicolumn{2}{c}{Rehit TTFT (ms)} & \multicolumn{2}{c}{Stored KV (MiB)} \\
\cmidrule(lr){3-4}\cmidrule(lr){5-6}\cmidrule(lr){7-8}
 & & Native & Ours & Native & Ours & Native & Ours \\
\midrule
Qwen & 32K & 4082.0 & 4204.3 & 547.3 & 307.1 & 1024.0 & 287.6 \\
 & 64K & 9105.3 & 9285.6 & 843.3 & 469.5 & 2048.0 & 577.1 \\
 & 128K & 22294.3 & 22491.2 & 1615.5 & 856.0 & 4096.0 & 1161.9 \\
\addlinespace[3pt]
GLM & 32K & 3268.9 & 3356.3 & 479.7 & 391.9 & 352.0 & 114.0 \\
 & 64K & 6390.0 & 6653.2 & 628.0 & 502.1 & 704.0 & 227.7 \\
 & 128K & 13077.2 & 14200.7 & 911.2 & 675.6 & 1408.0 & 454.7 \\
\addlinespace[3pt]
DeepSeek & 32K & 3414.0 & 3443.7 & 243.8 & 260.0 & 96.0 & 55.3 \\
 & 64K & 7176.2 & 7220.6 & 316.4 & 302.0 & 192.0 & 110.8 \\
 & 128K & 16228.4 & 16284.5 & 474.6 & 415.2 & 383.9 & 222.1 \\
\bottomrule
\end{tabular}

\end{table}
\subsection{Concurrent requests}
\label{sec:appendix-batch}
\begin{figure}[tbp]
\centering
\includegraphics[width=\linewidth]{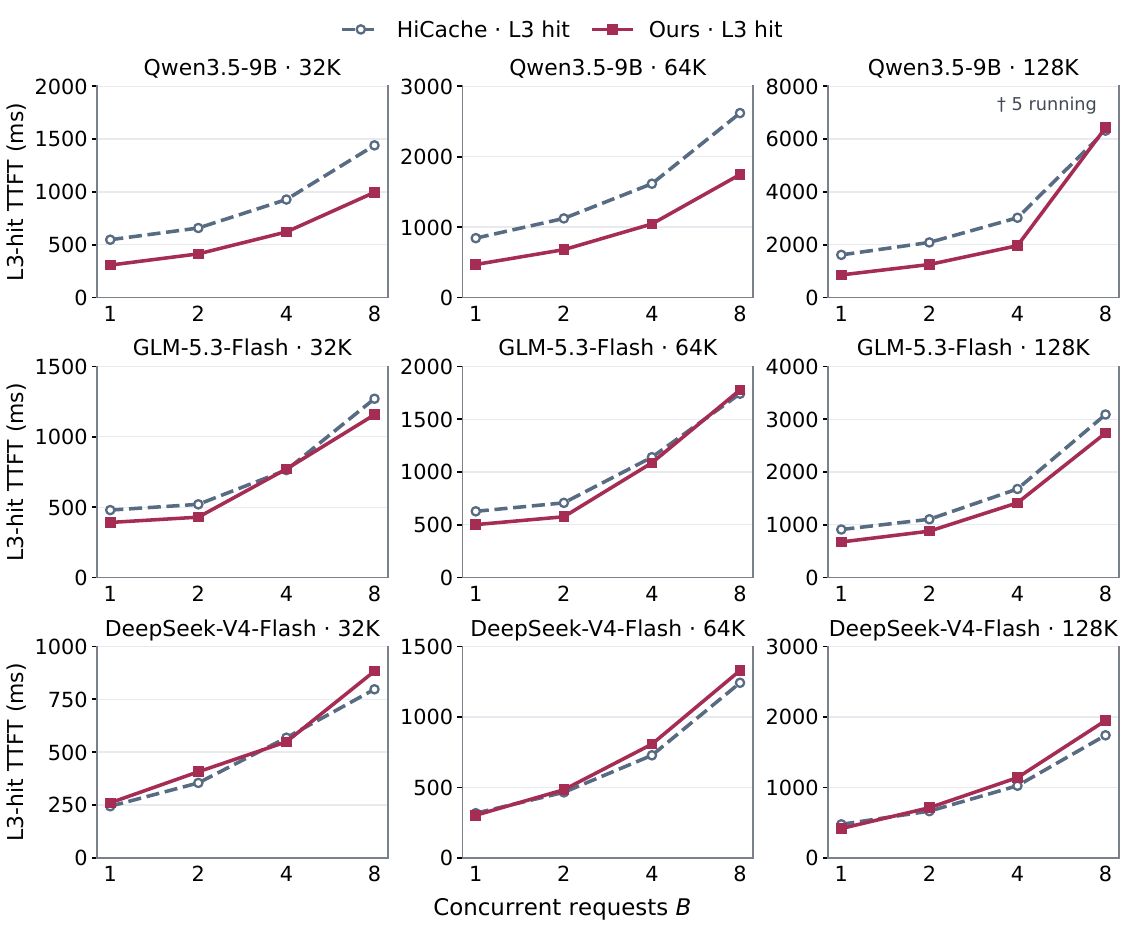}
\caption{Mean L3-hit restore TTFT of $B$ concurrent requests.
$\dagger$: GPU KV capacity admits only five requests in both arms. Axis
ranges vary by panel.}
\label{fig:serving-batch-ttft}
\end{figure}

Figure~\ref{fig:serving-batch-ttft} and Table~\ref{tab:batch-rehit-ttft}
report the mean L3-hit TTFT of $B$ concurrent requests.
With $B=2$--$8$, Qwen restores $31$--$40\%$ faster, except at 128K with
$B=8$. There, GPU KV capacity admits five of the eight requests in both arms,
and the other three wait until a running request finishes its 256 tokens.
Ours admits the first five requests earlier (mean TTFT $2.27$ versus
$3.52$\,s), so they decode together for longer and release a slot later.
GLM is faster at seven of nine points with $B\geq2$ and within $2\%$ at the
other two.
DeepSeek ranges from $3\%$ faster to $15\%$ slower: its C4 restoration runs
synchronously on the scheduler thread before each prefill, so consecutive
restores delay later requests.
Write TTFT with $B\geq2$ changes by $-0.7$ to $+6.5\%$ versus native HiCache.

\paragraph{Persistence completion.}
Table~\ref{tab:batch-rehit-ttft}(b) reports, for the same write requests, when
the batch's last target-KV page is written to L3. Both arms persist it
$1.5$--$6.5$\,s before the last generated token.

\begin{table}[tbp]
\centering\small
\setlength{\tabcolsep}{3pt}
\caption{$B$ concurrent requests. (a) Mean L3-hit restore TTFT (ms).
(b) Write-phase persistence completion (s), first submission to the L3 write
of the batch's last target-KV page. N: native HiCache, O: Ours $1\times$.
$\dagger$: GPU KV capacity admits only five requests in both arms.}
\label{tab:batch-rehit-ttft}
\textbf{(a)} Restore TTFT (ms)\par\vspace{2pt}
\begin{tabular}{llrrrrrrrr}
\toprule
Model & Ctx & \multicolumn{2}{c}{$B=1$} & \multicolumn{2}{c}{$B=2$} & \multicolumn{2}{c}{$B=4$} & \multicolumn{2}{c}{$B=8$} \\
\cmidrule(lr){3-4}\cmidrule(lr){5-6}\cmidrule(lr){7-8}\cmidrule(lr){9-10}
 & & N & O & N & O & N & O & N & O \\
\midrule
Qwen & 32K & 547 & 307 & 658 & 413 & 928 & 621 & 1441 & 996 \\
 & 64K & 843 & 470 & 1123 & 680 & 1616 & 1045 & 2619 & 1747 \\
 & 128K & 1615 & 856 & 2086 & 1251 & 3021 & 1963 & 6320$^\dagger$ & 6431$^\dagger$ \\
\addlinespace[3pt]
GLM & 32K & 480 & 392 & 521 & 429 & 764 & 771 & 1270 & 1157 \\
 & 64K & 628 & 502 & 708 & 576 & 1140 & 1087 & 1741 & 1776 \\
 & 128K & 911 & 676 & 1105 & 880 & 1679 & 1416 & 3089 & 2737 \\
\addlinespace[3pt]
DeepSeek & 32K & 244 & 260 & 354 & 407 & 568 & 549 & 797 & 883 \\
 & 64K & 316 & 302 & 464 & 483 & 727 & 807 & 1242 & 1329 \\
 & 128K & 475 & 415 & 662 & 711 & 1023 & 1138 & 1739 & 1948 \\
\bottomrule
\end{tabular}

\par\medskip
\textbf{(b)} Persistence completion (s)\par\vspace{2pt}
\begin{tabular}{llrrrrrrrr}
\toprule
Model & Ctx & \multicolumn{2}{c}{$B=1$} & \multicolumn{2}{c}{$B=2$} & \multicolumn{2}{c}{$B=4$} & \multicolumn{2}{c}{$B=8$} \\
\cmidrule(lr){3-4}\cmidrule(lr){5-6}\cmidrule(lr){7-8}\cmidrule(lr){9-10}
 & & N & O & N & O & N & O & N & O \\
\midrule
Qwen & 32K & 3.16 & 4.59 & 7.19 & 8.67 & 19.85 & 21.65 & 38.27 & 40.26 \\
 & 64K & 7.93 & 9.66 & 16.96 & 18.80 & 40.45 & 42.70 & 81.40 & 84.00 \\
 & 128K & 20.60 & 22.87 & 42.84 & 45.28 & 94.35 & 97.22 & 189.58 & 193.25 \\
\addlinespace[3pt]
GLM & 32K & 3.16 & 3.28 & 6.20 & 6.42 & 15.49 & 15.85 & 27.81 & 29.31 \\
 & 64K & 6.28 & 6.57 & 12.48 & 12.78 & 28.18 & 29.05 & 53.86 & 55.12 \\
 & 128K & 12.96 & 14.11 & 26.79 & 27.35 & 54.99 & 56.27 & 110.49 & 109.56 \\
\addlinespace[3pt]
DeepSeek & 32K & 3.29 & 3.38 & 6.57 & 6.77 & 13.10 & 13.21 & 25.94 & 26.09 \\
 & 64K & 7.05 & 7.15 & 14.07 & 14.28 & 28.08 & 28.26 & 55.85 & 56.07 \\
 & 128K & 16.10 & 16.19 & 32.08 & 32.37 & 64.16 & 64.41 & 128.15 & 128.51 \\
\bottomrule
\end{tabular}

\end{table}

\subsection{Coding-time measurement}
\label{sec:cpu-codec-throughput}
CPU encode and decode are timed on an AMD EPYC 9K84 with 32 workers and
host-resident inputs. We select 32 evenly spaced complete 256-position
objects from each of five documents (160 total), using the operating points
of Table~\ref{tab:matched-storage} and the same fitted parameters. Report
the median of three passes after one warmup. Timing includes layout
changes, quantization or dequantization, entropy coding or decoding,
allocation and packet handling, but excludes source loading, fitting and
validation.
The 160 objects hold 40,960 cache positions. A 32K-token input occupies
32,768 positions on Qwen and GLM but 8,192 C4 positions on DeepSeek, whose
C4 pool keeps one position per four input tokens
(Section~\ref{sec:calibration-details}). Scaling the encode times linearly
to these counts gives 1.012/0.540/0.195\,s per 32K input on Qwen/GLM/DeepSeek,
versus 4.082/3.269/3.414\,s native write TTFT at 32K with $B=1$
(Table~\ref{tab:online-roundtrip-complete}), a 4.0/6.0/17.5$\times$ throughput
headroom. L3-hit restoration measurements disable new L3 writes;
first-request measurements enable online writes.
\begin{table}[tbp]
\centering
\small
\caption{Same-host CPU encode and decode time (ms) at the
Table~\ref{tab:matched-storage} operating points (median of three passes
over 160 objects).}
\label{tab:matched-encode}
\begin{tabular}{llrrrr}
\toprule
Model & & Ours & KVCodec & KVTC-style & Rotation + SZ3 \\
\midrule
Qwen3.5-9B & Encode & \textbf{1266} & 7468 & 2079 & 1308 \\
 & Decode & \textbf{564} & 3321 & 1918 & 926 \\
\addlinespace
GLM-5.3-Flash & Encode & \textbf{676} & 3589 & 888 & 704 \\
 & Decode & \textbf{231} & 1239 & 759 & 493 \\
\addlinespace
DeepSeek-V4-Flash & Encode & \textbf{976} & 6675 & 1401 & 994 \\
 & Decode & \textbf{338} & 2126 & 1245 & 682 \\
\bottomrule
\end{tabular}
\end{table}

\FloatBarrier
\section{API Cache-Retention Policies}
\label{sec:api-cache-policies}
Figure~\ref{fig:api-cache-retention} summarizes public policies accessed on
September 20, 2026. Gemini defaults to 1 h~\citep{google2026cachepolicy}. OpenAI
supports up to 24 h on supported models~\citep{openai2026cachepolicy}. Baidu
allows 1 h--7 days~\citep{baidu2026cachepolicy}. DeepSeek's latest recorded
retention is 72 h~\citep{deepseek2026cachepolicy}.
These bars report the latest published retention for each provider.

\end{document}